\documentclass{jfm}
\usepackage{graphicx}
\usepackage{amssymb,bm}
\usepackage{xcolor}
\usepackage{amsmath}
\usepackage{siunitx}
\usepackage{pdflscape}
\usepackage{rotating}
\usepackage[export]{adjustbox}
\usepackage{lipsum}
\usepackage{multirow}
\usepackage{appendix}

\title{Dynamics of an inverted cantilever plate at moderate angle of attack}
\shorttitle{Dynamics of an inverted cantilever plate at moderate angle of attack}

\shortauthor{C. Huertas-Cerdeira, A. Goza, J.E. Sader, T. Colonius and M. Gharib}

\author{Cecilia Huertas-Cerdeira\aff{1}
  \corresp{\email{chuertas@caltech.edu}},
  Andres Goza\aff{2},
  John E. Sader\aff{3,4},
  Tim Colonius\aff{1}
 \and  Morteza Gharib\aff{1}}

\affiliation{\aff{1} Division of Engineering and Applied Science, California Institute of Technology, Pasadena, CA 91125, USA
\aff{2}Department of Aerospace Engineering, University of Illinois at Urbana-Champaign, Urbana, IL 61801, USA
\aff{3}ARC Centre of Excellence in Exciton Science, School of Mathematics and Statistics, The University of Melbourne, Victoria 3010, Australia
\aff{4}Department of Physics, California Institute of Technology, Pasadena, CA 91125, USA
}

\begin{document}

\maketitle

\begin{abstract}
The dynamics of a cantilever plate clamped at its trailing edge and placed at a moderate angle ($\alpha \leq \ang{30}$) to a uniform flow are investigated experimentally and numerically, and a large experimental data set is provided. The dynamics are shown to differ significantly from  the zero-angle-of-attack case, commonly called the inverted-flag configuration. Four distinct dynamical regimes arise at finite angles: a small oscillation around a small-deflection equilibrium (deformed regime), a small-amplitude flapping motion, a large-amplitude flapping motion and a small oscillation around a large-deflection equilibrium (deflected regime). The small-amplitude flapping motion appears gradually as the flow speed is increased and is consistent with a limit-cycle oscillation caused by the quasi-steady fluid forcing. The large-amplitude flapping motion is observed to appear at a constant critical flow speed that is independent of angle of attack. Its characteristics match those of the large-amplitude vortex-induced vibration present at zero angle of attack. The flow speed at which the plate enters the deflected regime decreases linearly as the angle of attack is increased, causing the flapping motion to disappear for angles of attack greater than $\alpha \approx \ang{28}$. Finally, the effect of aspect ratio on the plate dynamics is considered, with reduced aspect ratio plates being shown to lack sharp distinctions between regimes.

\end{abstract}

\section{Introduction} \label{sec:intro}

Elastic plates are in contact with flows in many natural settings and engineering designs. The interaction between the plate and the impinging flow often results in coupled fluid-structure physics and complex behaviors, with the resulting phenomena being dependent on the boundary conditions of the plate. Of particular interest is the behavior of elastic plates that are clamped on one edge and free to move on the remaining edges. This configuration is ubiquitous in nature; examples are sessile systems such as leaves \citep{Vogel1989} and elements of animal locomotion such as fish fins \citep{Sfakiotakis1999}. It is also present in many man-made systems, such as flags clamped to their pole \citep{Shelley2011} and the leaflets of artificial heart valves \citep{Driessen2007}. As is apparent from these examples, the direction at which the flow impinges on the plate plays a crucial role in the behavior of the system. 

Numerous studies have examined the behavior of elastic cantilever plates subjected to a uniform flow directed either perpendicular or parallel to the plate \citep{Shelley2011,Eloy2008,Paidoussis1998,Zhang2000,Vogel1994,Luhar2011,Kim2013}. In the case of flow perpendicular to the plate, the plate behaves as a bluff body and the main force acting on it is drag, together with any unsteady forces that may arise. These forces can produce large bending deformations that often result in a more streamlined shape, reducing in turn the drag force that the flow exerts on the plate \citep{Vogel1994}. This phenomenon, named reconfiguration, has been widely studied and is commonly seen in vegetation \citep{Delangre2008}. Conversely, when the plate is at small angles of attack to the flow, the flow remains attached and the force that acts perpendicularly to the plate --- which is responsible for its bending --- is due to lift. The resulting deflection of the plate can generate flow detachment and unsteady forces.

In the case of flow parallel to the cantilever plate, the flow can impinge either at the clamped or free end of the plate. The former case is the most frequently studied, and is referred to as the conventional flag configuration because of its similarity to a flag flapping in the wind. For this flag system, the flag's mass and flexibility act as bifurcation parameters; for certain ranges of these parameters, the flag can exhibit flapping about its undeflected equilibrium through a flutter instability mechanism \citep{Shelley2011}. Both limit cycle and chaotic flapping modes have been observed \citep{Alben2008}. Conventional flags have been proposed as a means to harvest energy from the wind through the use of piezoelectric materials, which convert the strain energy induced by the flapping motion into electric energy \citep{Allen2001,Taylor2001,Lee2015}. Recent studies, however, have focused increasingly on the former case---the inverted flag---where the flow is parallel to the undeflected plate and impinges on its free edge. The large amplitudes of motion and strains exhibited by the inverted flag make this configuration particularly useful for applications that involve energy harvesting, mixing and heat transfer enhancement \citep{Kim2013,Orrego2017,Park2016}. 

\cite{Kim2013} showed that the inverted flag presents three main regimes of motion depending on the magnitude of the dimensionless flow speed $\kappa = \rho U^2 L^3 / D$, where $\rho$ is the fluid density, $U$ is the free stream flow speed, $L$ the length of the flag and $D$ the flexural rigidity of the flag. At low flow speeds, the flag remains in the straight regime, characterized by its zero deflection equilibrium. The flow remains attached in this regime \citep{Gurugubelli2015,Goza2018}. As wind speed is increased, the system becomes unstable and enters the large-amplitude flapping regime. The shedding frequency of vortex structures in this regime is correlated to this flapping motion, with a variety of vortex patterns occurring for different flow speeds \citep{Kim2013,Ryu2015,Gurugubelli2015,Shoele2016,Goza2018}. If the wind speed is further increased, the flag enters the deflected regime, where it oscillates with small amplitude around a deflected equilibrium. Bi-stable regions are present both in the transition from straight to flapping and from flapping to deflected regimes \citep{Kim2013}, with a small region of chaotic flapping present between the flapping and deflected regimes \citep{Sader2016a}. Computational studies have reported the existence of additional regimes. \cite{Ryu2015} and \cite{Gurugubelli2015} observed both a small-deflection steady state and a small-deflection small-amplitude flapping regime at flow speeds between those corresponding to the straight and large amplitude flapping regimes. \cite{Goza2018} demonstrated the small-deflection small-amplitude flapping to be a supercritical Hopf bifurcation of the small-deflection equilibrium state. \cite{Gurugubelli2015},\cite{Tang2015} and \cite{Shoele2016} additionally observed a flipped flapping regime at wind speeds higher than those of the deflected regime. In this mode, the flag bends \ang{180}  such that the leading edge is parallel to the flow, recovering a motion similar to that of the conventional flag. Overall, the regimes of motion that have been reported for the inverted flag are, ordered from lowest to highest corresponding flow speed: straight, small-deflection steady, small-deflection small-amplitude flapping, large amplitude flapping, chaotic, deflected and flipped flapping.

In an attempt to understand the onset of the large amplitude flapping motion of the inverted flag, several studies have investigated the loss of stability of the straight regime. While the existence of a divergence instability was hinted by \cite{Kim2013}, \cite{Gurugubelli2015} was the first to numerically demonstrate its presence. \cite{Sader2016a} theoretically corroborated the loss of stability of the straight regime through divergence, and provided a simplified analytic formula that reasonably predicts the onset of flapping for inverted flags of aspect ratios higher than 1. An equivalent formula, valid for varying flag morphologies, was further developed by \cite{Fan2019}. The inverted flag's flapping motion was shown by \cite{Sader2016a} to constitute a vortex-induced vibration (VIV). \cite{Goza2018} associated this classic vortex-induced vibration with the 2P vortex shedding mode, and linked the appearance of additional shedding modes at higher flow speeds with the breakdown of the VIV and appearance of chaos. The cessation of flapping has received comparably little attention, and is not yet fully understood. \cite{Sader2016a} suggested the mechanism behind this transition to be a disruption of lock-on caused by the increased disparity between natural and shedding frequencies of the flag. 

It should be noted that the above described behavior is valid for inverted flags of large aspect ratio only. As aspect ratio is decreased the nonlinear fluid loading induced by the edge vortices significantly modifies the flag's dynamics. The increased lift force results in a decrease of the flow speed at which divergence occurs. At small aspect ratios, \cite{Sader2016b} showed that inverted flags undergo a saddle-node bifurcation, which occurs at flow speeds lower than those predicted for a divergence instability. Notably, for $AR<0.2$ the large-amplitude flapping motion disappears \citep{Sader2016a}. The conclusions about large-amplitude flapping in large-aspect-ratio flags have, additionally, been drawn in the case of heavy fluid loading, in which the ratio of flag to flow inertia is small. \cite{Goza2018} demonstrated that large-amplitude flapping persists in the presence of light fluid loading, but is distinct from classical VIV. Large-amplitude flapping occurs for very heavy flags at the low Reynolds number of $20$ ---well below the critical Reynolds number of $\approx47$ for which bluff-body vortex shedding initiates in the case of a circular cylinder.  

Laboratory and numerical studies have highlighted the energy harvesting potential of the inverted flag \citep{Kim2013,Ryu2015,Shoele2016,Silva2019}. Field realizations of the energy harvester have shown, however, that the frequent changes in flow direction characteristic of atmospheric winds result in reduced harvesting performance \citep{Orrego2017}. Changes in flow direction correspond to variations in the angle of attack of the undeflected flag, which modify its flapping dynamics. This undeflected angle of attack is equal to the angle between the clamping direction of the trailing edge and the direction of the impinging flow (figure \ref{setup}b). It will be referred to as \emph{angle of attack}, $\alpha$, throughout this text and is independent of the flag's motion. It should be noted that the instantaneous angle of attack of the flag, that is time and position dependent as the flag deflects, is not used in this text. The dynamics of the inverted flag are very susceptible to changes in the angle of attack, $\alpha$, as can be deduced from the very different behaviors in the $\alpha=\ang{90}$ (reconfiguration) and $\alpha=\ang{180}$ (regular flag) limits described above. Although these two limiting cases have been studied thoroughly, very little information is known about the behavior of cantilevered plates at intermediate angles of attack.

Preliminary wind tunnel tests have been performed by \cite{Cosse2014} on an inverted flag of aspect ratio $AR=2$ at angles of attack of $\alpha=\ang{0}$, $\ang{10}$  and $\ang{20}$. At finite $\alpha$, the plate exhibits a gradual increase in its amplitude of motion as the flow speed ($\kappa$) is increased. This behavior is distinct from the zero-angle-of-attack case, $\alpha=\ang{0}$, where the plate presents an abrupt increase in amplitude at a distinct value of $\kappa$. The critical $\kappa$ at which the plate transitions from the flapping to the deflected regime is different for the three angles. Additionally, the plate at an angle of $\alpha=\ang{20}$ showed smaller maximum flapping amplitudes than those at smaller angles. This has been corroborated by the brief computational studies of \cite{Shoele2016} and \cite{Tang2015}, who observed that the maximum flapping amplitude decreases abruptly for angles of attack larger than $\alpha=\ang{15}$ degrees. Recently, \cite{Tavallaeinejad2020a} and \cite{Tavallaeinejad2020b} developed a non-linear analytical model of small-aspect-ratio and two-dimensional inverted flags, respectively, and investigated the sensitivity of the model to angle of attack. In the two-dimensional model, the flag presents a gradual increase in amplitude as flow speed is increased, in agreement with the result obtained by \cite{Cosse2014}, with the critical flow speed decreasing as angle of attack is increased. The flag then transitions to a large-amplitude flapping motion through two saddle-node bifurcations, marked by a jump in the frequency of motion. In the low aspect ratio model, the pitchfork bifurcation present at zero angle of attack is replaced by a saddle-node bifurcation at non-zero angles. At large angles, this saddle-node bifurcation produces an increased region of bi-stability. 

While these results highlight fundamental changes in the inverted flag's dynamics with angle of attack, the literature lacks a comprehensive characterization of the behavior of inverted cantilever plates at varying angles. The term ``inverted cantilever'' will be used throughout this text as a more generalized denomination for inverted flags. It refers to a cantilevered elastic plate that is clamped at an angle to the impinging flow such that its leading edge corresponds to the free edge of the plate. The purpose of this study is to fully characterize the dynamics of inverted cantilevers for angles of attack between $\alpha=\ang{0}$ and $\alpha=\ang{30}$, and in so doing, to generate a more comprehensive experimental data set of the system. Results are primarily drawn from a collection of systematically performed experiments at high Reynolds numbers ($Re \approx 10^4$, see section \ref{setup} for definitions). In addition, targeted two-dimensional, low Reynolds number ($Re=200$) simulations are employed to demonstrate the robustness of the behavior across Reynolds number and to clarify physical mechanisms driving certain observed behaviors. Angles of attack $\alpha \leq \ang{30}$ are considered because, as will be shown, the large-amplitude flapping motion disappears beyond that value for plates of moderately large aspect ratios. 

The experimental setup employed for this study, as well as the non-dimensional parameters relevant to the characterization of the plate's dynamics, are introduced in Section \ref{sec:setup}. The numerical method used to obtain computational results is presented in Section \ref{sec:NumMethod}. The primary results of the article are provided in sections \ref{AoA_0}--\ref{AR2}. First, we review and extend conclusions about an inverted flag at $\alpha=\ang{0}$ that form the foundation of the current work (section \ref{AoA_0}). Next, the system dynamics at nonzero angle of attack (i.e. $\alpha>\ang{0}$) are divided into four different regimes, and the physical mechanisms behind each regime are clarified (section \ref{AR5}). The evolution of these regimes as the angle of attack is increased is then synthesized in section \ref{AoA_evolution}. Finally, in section \ref{AR2} the effect of aspect ratio is investigated by comparing the results from sections \ref{AoA_0}--\ref{AoA_evolution} to those for a plate of lower aspect ratio ($AR=2$). The findings of this article are summarized in section \ref{conclusions}.

 \section{Experimental setup} \label{sec:setup}

\begin{figure}
         \centerline{\includegraphics[width=120mm]{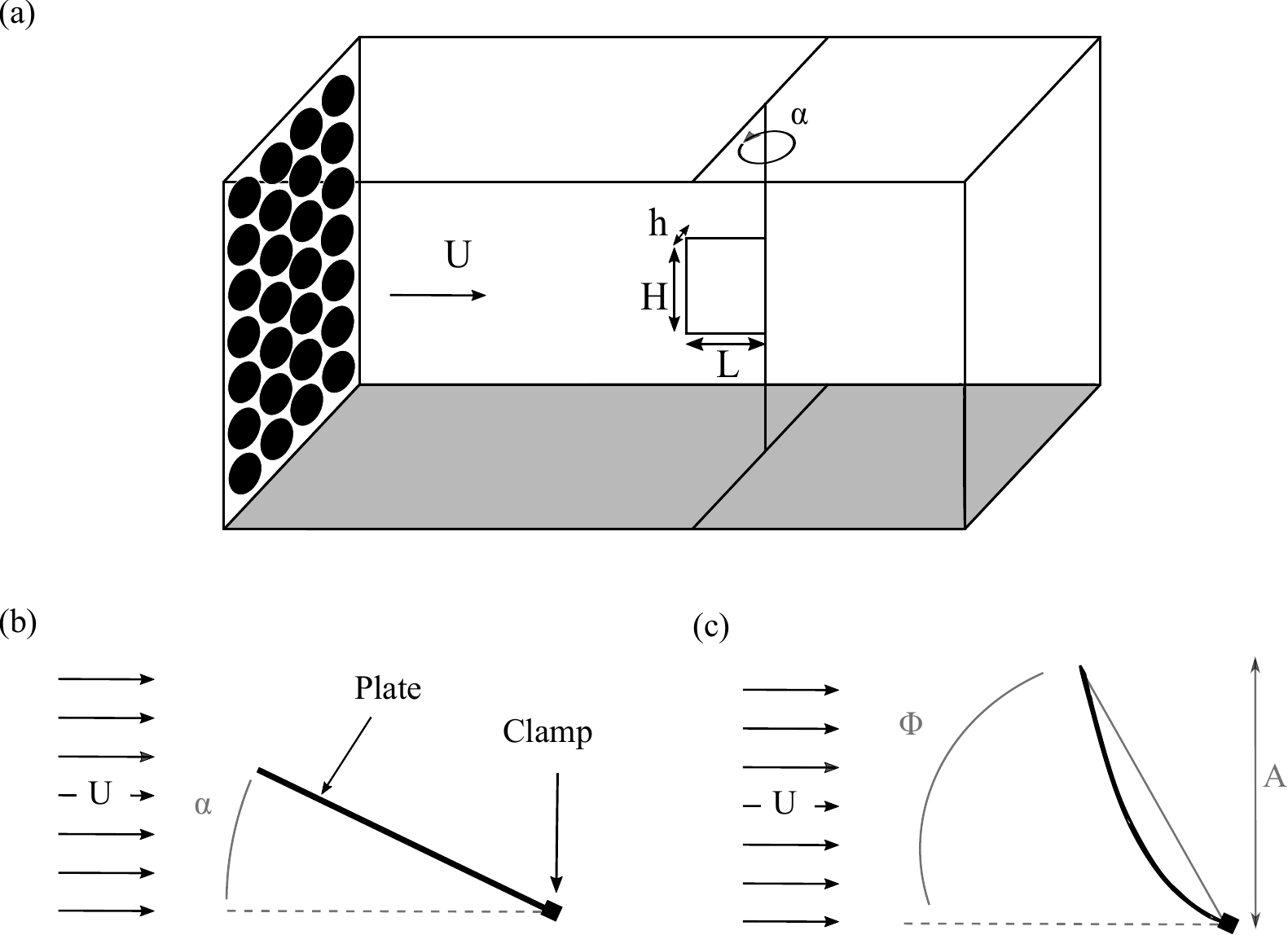}}
        \caption{(a) Schematic of the experimental setup, with notation for plate dimensions. Top view of a deflected inverted cantilever plate with definition of (b) angle of attack, $\alpha$, and (c) deflection angle, $\Phi$, and amplitude, A.}
        \label{setup}
\end{figure}

The experiments are conducted in an open-loop wind tunnel with a test section of cross section $1.2m \times 1.2m$. A schematic of the setup is presented in figure \ref{setup}a. The flow is generated by an array of $10\times10$ small fans that can produce uniform flow speeds between 2.2 and 8.5 m/s, with a maximum turbulence intensity of 9.8\% for the range of wind speeds considered in this study. The plates are made of polycarbonate (Young's modulus $E=2.41 \ \si{GPa}$, Poisson ratio $\nu=0.38$, density $\rho_s=1200 \ \si{kg.m^{-3}}$ ). The main analysis is performed on a plate of aspect ratio 5, as defined in equation \ref{eq:parameters}, which has a length of $L=82\ \si{mm}$, height of $H=410\ \si{mm}$ and  thickness of $h=0.254\ \si{mm}$ (figure \ref{setup}a).  A plate of aspect ratio 2 and dimensions $L=160\ \si{mm}$, $H=320\ \si{mm}$ and $h=0.508\ \si{mm}$ is subsequently investigated to account for the variability of the results with aspect ratio. The plates are clamped vertically at their trailing edge by means of two aluminium bars, which have rectangular cross section of width 6mm x 12mm and height equal to that of the test section ($1.2\ \si{m}$). The bars are mounted on an adjustable hinge that allows a predetermined and fixed rotation along the vertical axis, corresponding to different values of the angle of attack of the undeflected plate $\alpha$ (figure \ref{setup}b). The value of the rotation is set through a dial that displays $\ang{2}$ increments. 

The deformation of the plate is primarily two-dimensional in the horizontal plane, with no twisting due to gravity being observed. Its motion is recorded with a high speed camera (Imperx IPX-VGA210-L) located above the test section. Two recordings of each motion are obtained, one 600 frames long at 20fps, used for frequency content analysis, and one 200 frames long at 100fps, used for the remaining data. The coordinates of the top edge of the plate are extracted for each frame through an edge recognition algorithm implemented in Matlab. 
In addition to the angle of attack, $\alpha$, there are four non-dimensional parameters that define the dynamics of an inverted cantilevered plate: Reynolds number, $Re$, non-dimensional speed, $\kappa$, aspect ratio, $\mathrm{AR}$, and mass ratio, $\mu$. They are defined as follows

\begin{equation}\label{eq:parameters}
Re=\frac{U L}{\nu},
\qquad
\kappa= \frac{\rho_f U^2 L^3}{D} ,
\qquad
\mathrm{AR}=\frac{H}{L} ,
\qquad
\mu=\frac{\rho_s h}{\rho_f L} 
\end{equation}

\noindent with $U$ the free stream flow speed, $\nu$ the kinematic viscosity of the fluid, $\rho_s$ the density of the plate, $D$ the flexural rigidity of the plate and $\rho_f$ the density of the fluid. Due to the constant dimensions, the mass ratio of each plate remains constant throughout the experiments, with a value of of  $\mu=3.0$ for the plate of $\mathrm{AR}=5$ and $\mu=3.1$ for the plate of $\mathrm{AR}=2$. The flow speed is varied to obtain data for different values of $\kappa$, yielding a Reynolds number that varies between $Re=1.4-4.4 \times 10^4$.

The resulting motion of the plate is characterized throughout this text using four main parameters: the deflection angle, $\Phi$, the non-dimensional amplitude, $\Tilde{A}$, provided for comparison with previous studies, the non-dimensional frequency, $\Tilde{f}$ and the Strouhal number, $St$. The deflection angle, $\Phi$, is represented in figure \ref{setup}c and corresponds to the angle between the free stream direction and the line joining the leading and trailing edges of the plate. The sign of this angle is determined by the side to which the plate deflects, with the positive direction being upwards in this figure. Variables derived from this parameter, such as the angular amplitude $\Delta \Phi= \Phi_{max}-\Phi_{min}$ and the mean deflection angle, $\Bar{\Phi}$, will be employed occasionally. The non-dimensional amplitude, non-dimensional frequency and Strouhal number are defined as

\begin{equation*}
\Tilde{A}=\frac{A}{L},
\qquad
\Tilde{f}=\frac{f L}{U},
\qquad
St=\frac{f A'}{U},
\end{equation*}

\noindent where $A$ is the signed amplitude of motion as defined in figure \ref{setup}c, with its sign being positive in the upwards direction, and $f$ is the frequency of oscillation. The cross section $A'$ is calculated as the maximum between $A_{max}-A_{min}$ and $|A_{max}|$.

\section{Numerical Method}  \label{sec:NumMethod}
 
In order to clarify some of the underlying physics behind the experimentally observed plate behavior, two-dimensional numerical simulations are performed for an inverted cantilevered plate at an angle of attack of $\alpha=\ang{10}$. 

The nonlinear simulations are performed using the immersed-boundary algorithm of \citet{Goza2017} for $Re=200$ and $\mu=0.5$. The method treats the fluid equations using a discrete streamfunction formulation \citep{Colonius2008} and the plate (modeled as a geometrically nonlinear Euler-Bernoulli beam) with a corotational finite element formulation \citep{Crisfield1991}. The method is strongly coupled; \emph{i.e.}, the nonlinear coupling between the structure and the fluid is enforced at each time instance, resulting in a stable algorithm in the presence of large displacements and rotations. This fluid-structure coupling is enforced by the stresses on the immersed surface, and immersed-boundary methods are well known to produce spurious computations of these stresses. These unphysical stress computations were remedied by \citet{Goza2016}, and the techniques described therein are incorporated into the FSI algorithm of \citet{Goza2017} to ensure appropriate treatment of the fluid-structure coupling. The FSI solver has previously been validated on several flapping plate problems for plates in both the conventional configuration (pinned or clamped at the leading edge) and the inverted configuration (clamped at the trailing edge) in \cite{Goza2017}.

The flow equations are treated using a multidomain approach: the finest grid surrounds the body and grids of increasing coarseness are used at progressively larger distances \citep{Colonius2008}. All spatial dimensions are scaled by the cantilever length, $L$, whereas time is non-dimensionalized using $L/U$. The cartesian x-axis is in the flow direction and the y-axis points vertically upwards; the z-axis is not required in these 2D simulations. In all computations below, the domain size of the finest sub-domain is $[-0.2, 1.8] \times [-1.1, 1.1]$ and the total domain size is $[-15.04, 16.64] \times [-17.44, 17.44]$. The grid spacing on the finest domain is $\Delta x=\Delta y = 0.01$ and the grid spacing along the flag is $\Delta s = 0.02$. The time step is $\Delta t = 0.001$, which gives a maximum Courant-Friedrichs-Levy number of $\approx 0.15$. These simulation parameters are identical to what was used in \citet{Goza2018}, and Appendix A of that reference demonstrates the suitability of these parameters for the Reynolds number considered in the numerical portion of the present study. 

\section{Dynamics at zero angle of attack}\label{AoA_0}

\begin{figure}
	\includegraphics[width=\textwidth]{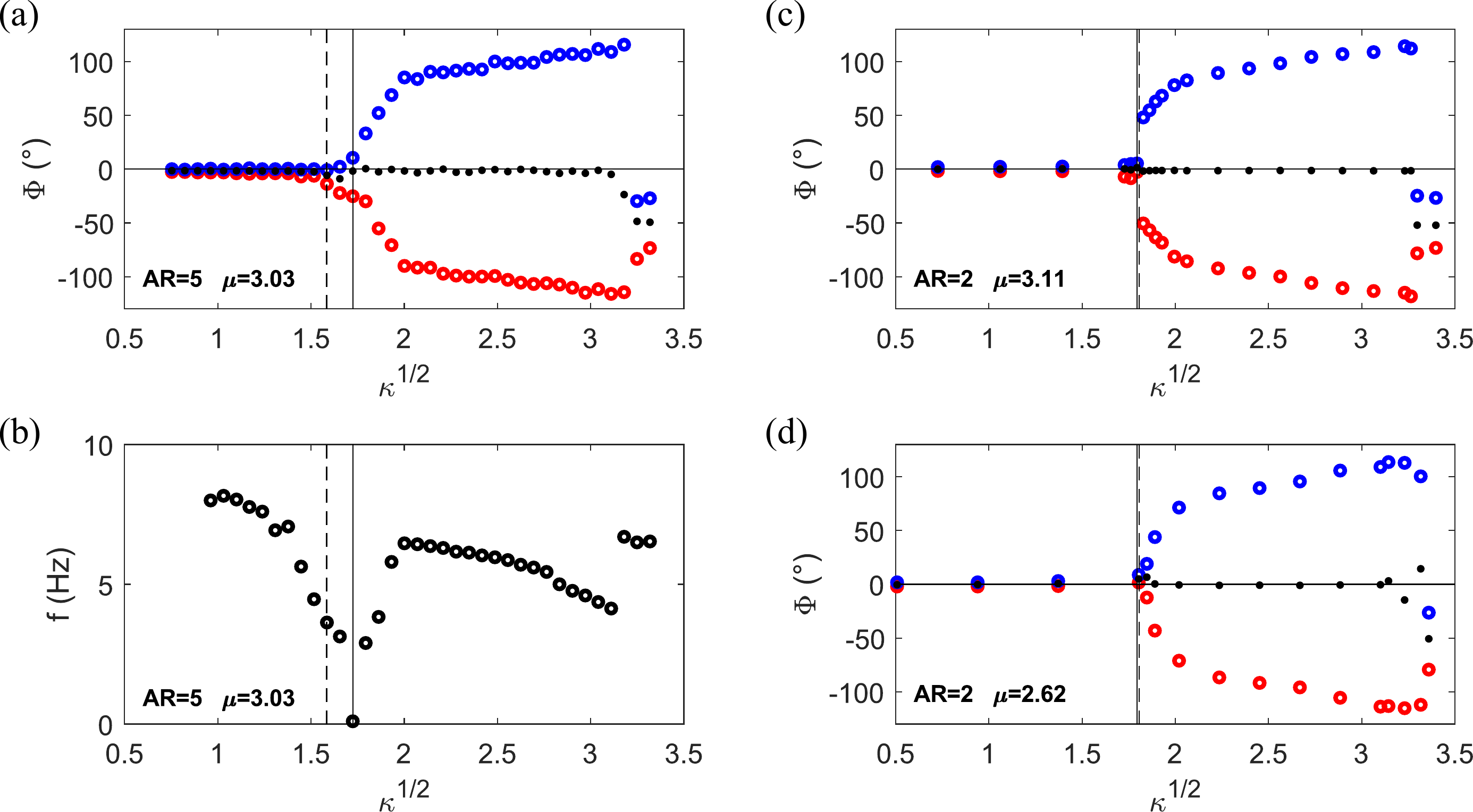}
        \caption{Behavior of an inverted cantilever plate at zero angle of attack, obtained experimentally. Maximum (\textcolor{blue}{$\circ$}), minimum (\textcolor{red}{$\circ$})  and mean ($\bullet$) deflection angle, $\Phi$, for a plate of (a) AR=5 and $\mu=3.03$, (c) AR=2 and $\mu=3.11$  and (d) AR=2 and $\mu=2.62$. (b) Frequency of motion of the plate of AR=5 and $\mu=3.03$. Experimentally measured divergence flow speed (---) and theoretical prediction of divergence flow speed, using \cite{Sader2016b}(- -).}
        \label{alpha0}
\end{figure}

The experimental results obtained in this study at zero angle of attack are consistent with the existing literature. Figure \ref{alpha0}a shows the maximum, minimum and average deflection angle, $\Phi$, for a plate of AR=5 as a function of $\kappa^{1/2}$, which is proportional to the impinging flow speed (see Eq. \ref{eq:parameters}). The three main dynamical regimes \citep{Kim2013} are clearly visible. At low flow speeds, the plate remains undeflected, undergoing a small amplitude oscillation (straight regime). As wind speed is increased, the flag enters the large-amplitude flapping regime with a symmetric flapping motion of amplitude comparable to the flag length. For the highest values of $\kappa^{1/2}$, the plate enters the deflected regime and flexes to one side, oscillating with relatively small amplitude around the deflected position. 

As discussed above, the transition from straight to flapping regimes has been demonstrated numerically \citep{Gurugubelli2015} and theoretically \citep{Sader2016a} to be caused by a divergence instability of the zero deflection equilibrium. Figure \ref{alpha0}b constitutes the experimental verification of the existence of this divergence, which is absent in the existing literature. The figure displays the frequency of motion as a function of non-dimensional flow speed. The small amplitude oscillations of the plate in the straight regime are caused by small flow perturbations, and their frequency is equal to the plate's effective natural frequency, which includes the dampening effect of the fluid. In the presence of a divergence instability, the effective stiffness of the plate should asymptote to zero, with the natural frequency thus following the same trend. The trend can be clearly observed in figure \ref{alpha0}b. The value of the oscillation frequency at $\kappa^{1/2}=1.72$ is equal to $\Tilde{f}=0.1 \si{Hz}$, indicating the proximity of the divergence point. This experimental value of the divergence flow speed has been marked with a solid line in figures \ref{alpha0}a and \ref{alpha0}b. The theoretical value obtained using Eq. (2.15) in \cite{Sader2016b}, $\kappa^{1/2}=1.58$, is in reasonable agreement and is represented for reference with a dashed line.

As opposed to the results presented in previous experimental studies for zero angle of attack \citep{Kim2013,Huertas2018,Cosse2014}, the amplitude of motion in figure \ref{alpha0}a does not increase abruptly after the divergence instability. In order to demonstrate that this discrepancy in post-critical behavior is not a result of variations in either aspect ratio or mass ratio with respect to previous studies, the maximum, minimum and average deflection angles for plates of $\mu=3.11$ and $\mu=2.62$ and AR=2 are presented in figures \ref{alpha0}c and \ref{alpha0}d, respectively. The experimental divergence flow speed, $\kappa^{1/2}=1.80$, and theoretical divergence flow speed obtained from Eq.(2.15) of \cite{Sader2016b}, $\kappa^{1/2}=1.81$, of these plates are almost identical. Figure \ref{alpha0}c exhibits a discontinuous jump in amplitude of motion after the divergence, while figure \ref{alpha0}d shows a gradual increase. Because both plates correspond to the same aspect ratio, the differing behaviors cannot be caused by aspect ratio effects. Additionally, the mass ratio of all three plates is very similar, with the plate of $AR=5$ (figure \ref{alpha0}a, $\mu=3.03$) possessing a mass ratio that is closer to that of the plate presenting an abrupt transition (figure \ref{alpha0}c, $\mu=3.11$) than to that of the plate presenting a smooth transition (figure \ref{alpha0}d, $\mu=2.62$). The discrepancy in post-bifurcation behavior is, therefore, most likely due to small experimental differences, such as variations in the initial curvature of the plate.

It is interesting to note that the maximum amplitude of motion of the plate of $AR=5$ (figure \ref{alpha0}a) increases visibly as it approaches the divergence point from the lower flow speeds. This growth corresponds to an increase in the deflection of the plate when it is subject to perturbations: as the divergence flow speed is approached, the effective stiffness of the plate decreases and, therefore, any perturbation will generate larger amplitude deviations. These appear as broadband noise in the time history of the deflection angle $\Phi$ and do not correspond to a flapping motion.

\section{Dynamics at non-zero angle of attack} \label{AR5}

\subsection{Regimes}

The behavior of an inverted cantilever plate that is clamped at a non-zero angle of attack to the flow differs significantly from that of a plate placed at zero angle of attack, with new dynamical regimes emerging at non-zero angles of attack. To introduce these new dynamical regimes, we first present results for an angle $\alpha = \ang{10}$ as an illustrative example. 

\begin{figure}
	\includegraphics[width=\textwidth]{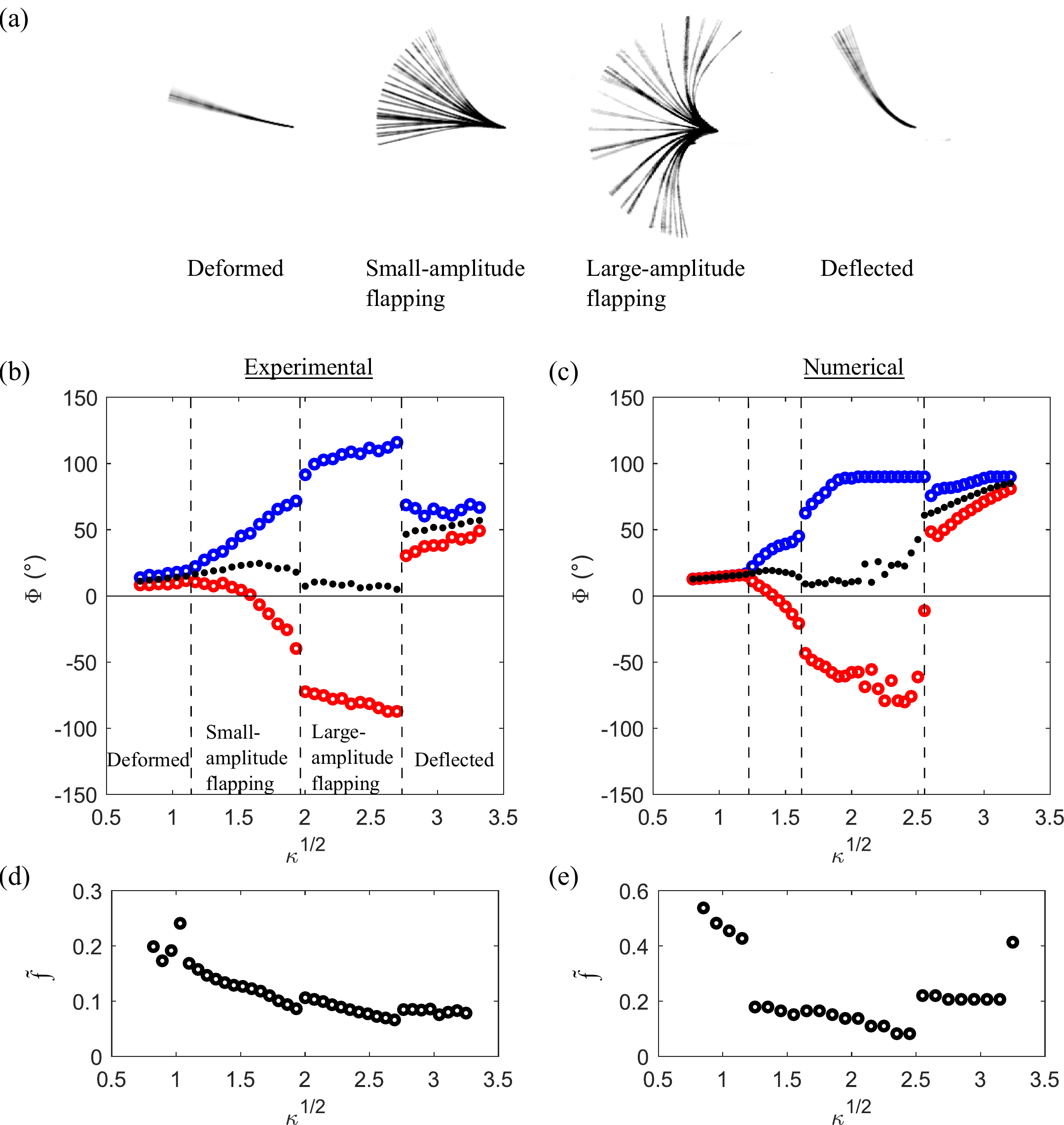}
        \caption{Behavior of an inverted cantilever plate at finite angle of attack ($\alpha=\ang{10}$). (a) Superimposed snapshots of the plate throughout its motion, depicting the four dynamical regimes. (b, c) Maximum (\textcolor{blue}{$\circ$}), minimum (\textcolor{red}{$\circ$}) and mean ($\bullet$) deflection angle, $\Phi$, as a function of non-dimensional flow speed, $\kappa$. (d, e)  Non-dimensional frequency of motion, $\Tilde{f}$, as a function of non-dimensional flow speed, $\kappa$. The results were obtained experimentally for a plate of AR=5, $\mu=3.0$ and $Re\sim\mathcal{O}(4)$ (a, b and d) and numerically for a two-dimensional inverted cantilever at $\mu=0.5,Re=200$. (c and e). The nomenclature employed for the different dynamical regimes is specified in (a, b). }
        \label{modes}
\end{figure}

\begin{figure}
	\includegraphics[width=\textwidth]{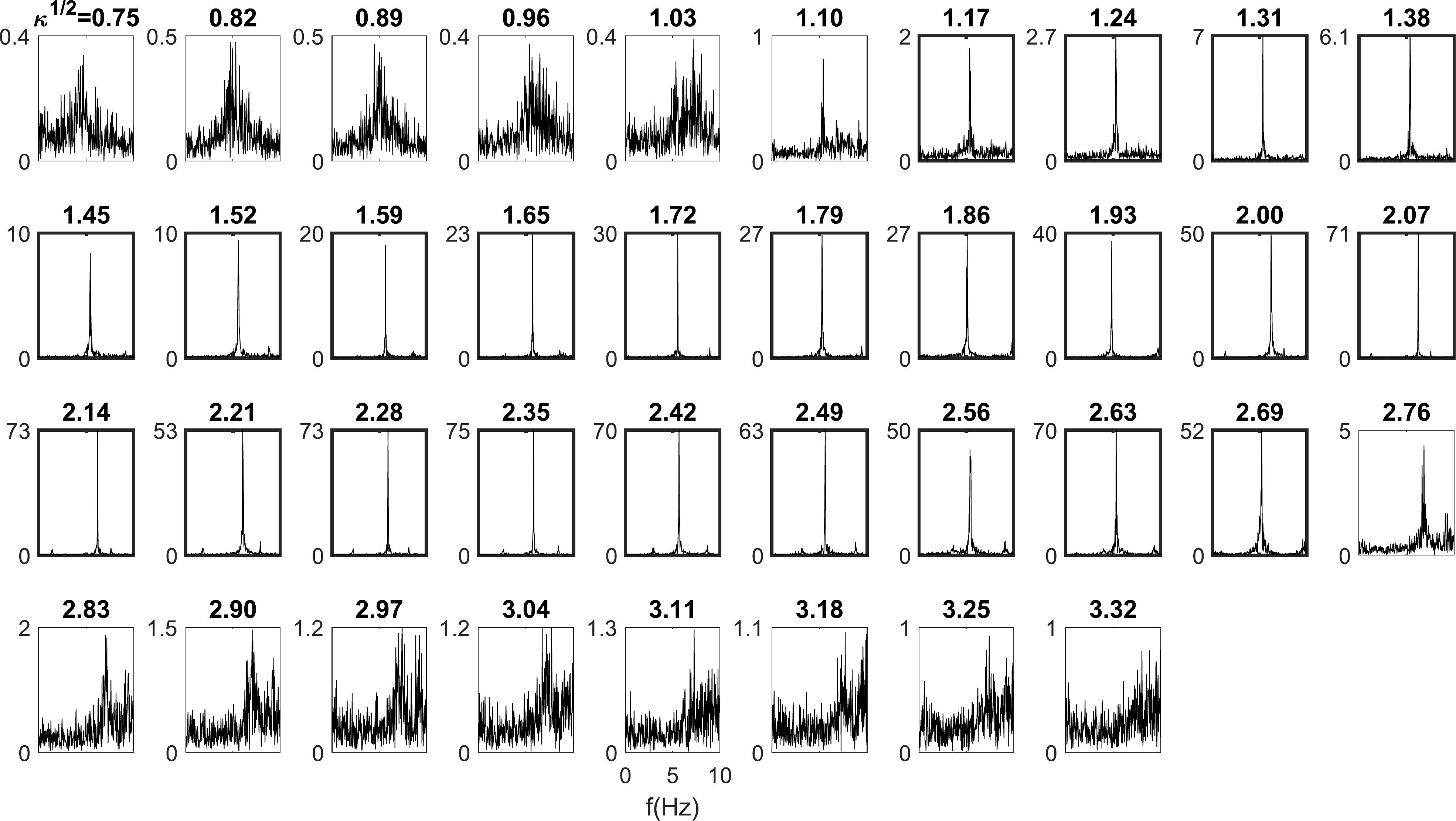}
        \caption{Power spectra of the inverted cantilever plate's deflection measured experimentally for $\alpha=\ang{10}$ and varying flow speeds. The flow speeds highlighted with a bold frame correspond to the flapping regime as defined by equation (\ref{flapcondition}). The scale varies and is specified for each spectrum.}        \label{fft10}
\end{figure}

\textit{Experimental measurements}. The four main dynamical regimes present at non-zero angle of attack are depicted in figure \ref{modes}a, which shows superimposed snapshots of the motion of a plate of $AR=5$ at $\alpha=\ang{10}$. Figure \ref{modes}b shows the maximum, minimum and average deflection angle, $\Phi$, measured experimentally for the same plate as a function of non-dimensional flow speed, $\kappa^{1/2}$. The dynamical regimes present in the plate's motion are demarcated in both figures. The \emph{deformed} regime occurs at low flow speeds. In it the plate oscillates with small amplitude around a small-deflection position. As wind speed is increased, these oscillations grow into a flapping motion. This flapping motion is divided into two branches, the \emph{small-amplitude flapping regime} and the \emph{large-amplitude flapping regime}, which will be shown in the following sections to constitute distinct dynamics. Finally, as wind speed is further increased the plate enters the \emph{deflected} regime, where it oscillates with small amplitude around a large-deflection position. The dominant non-dimensional frequency of oscillation of the plate, $\Tilde{f}$, is presented as a function of $\kappa^{1/2}$ in figure \ref{modes}d. It corresponds to the frequency at which the power spectrum of the flag's deflection presents its maximum peak. The spectrum is calculated using the fast Fourier transform of the deflection angle time series, $\Phi(t)$. As a reference, the spectra for the $\alpha=\ang{10}$ case are plotted in figure \ref{fft10}. Experimental frequency data throughout this text is presented only for flow speeds at which the spectra presents a clear peak. In some cases, two distinct peaks are present; the peak of largest amplitude is taken as the dominant frequency. Because their amplitudes are similar, small changes result in the switching of the dominant peak, which is reflected as a jump in the presented frequency. An example of such a jump is the data point for $\kappa^{1/2}=1.03$ in figure \ref{modes}d. A distinction should be made between these jumps in frequency and those present at the changes of dynamical regime, which correspond to a shift in the location of a single peak in the spectra.

\textit{Simulations}. The maximum, minimum and average deflection angles obtained numerically, as described in section \ref{sec:NumMethod}, for $Re=200$ and $\alpha=\ang{10}$, are displayed as a function of $\kappa^{1/2}$ in figure \ref{modes}c. Despite the proximity of the simulations to a bifurcation in $Re$, the numerical results exhibit qualitative agreement with the experiments and show the presence of the same four dynamical regimes. The flow speed at which the small-amplitude flapping emerges is in reasonable agreement ($\kappa^{1/2}=1.17$ for experiments, $\kappa^{1/2}=1.25$ for numerical simulations). However, the flow speed at which the large amplitude flapping starts differs significantly ($\kappa^{1/2}=2.00$ for experiments, $\kappa^{1/2}=1.65$ for numerical simulations) and the flow speed at which the large-amplitude flapping ceases sees moderate differences ($\kappa^{1/2}=2.76$ for experiments, $\kappa^{1/2}=2.6$ for numerical simulations). These differences are to be expected because the large-amplitude flapping motion is strongly dependent on the vortex shedding processes \citep{Sader2016a}, which vary with the Reynolds number. The frequencies of motion, $\Tilde{f}$, obtained numerically for the flag at $\alpha=\ang{10}$ are presented in figure \ref{modes}e. The trend and values within the flapping motion are in agreement between experimental and numerical results (notice the difference in scale), and the jumps in frequency at the start and end of the flapping regime are present in both cases, despite significant variations in the values at deformed and deflected regimes. 

Overall, the results obtained numerically and experimentally present appreciable differences but display the same general dynamical features. This is in agreement with results reported for an inverted cantilever plate at $\alpha=\ang{0}$, which show the characteristic qualitative features of the plate's dynamics and vortex wake to be fairly insensitive to Reynolds number for $Re>100$, while differences do occur in certain details as a function of $Re$ \citep{Tang2015,Shoele2016}. In this study, the majority of the results will be extracted from the experimental data, and the numerical results will be employed to gain a more detailed understanding of i) the plate behavior in the flapping regime, and ii) the difference in flow structures between the small- and large-amplitude flapping regimes. Despite the non-negligible differences, the conclusions extracted from the numerical data in this text refer to the common general physics and are expected to hold for the large $Re$ case.

\begin{landscape}
\begin{figure}
	\includegraphics[height=0.78\textwidth,keepaspectratio]{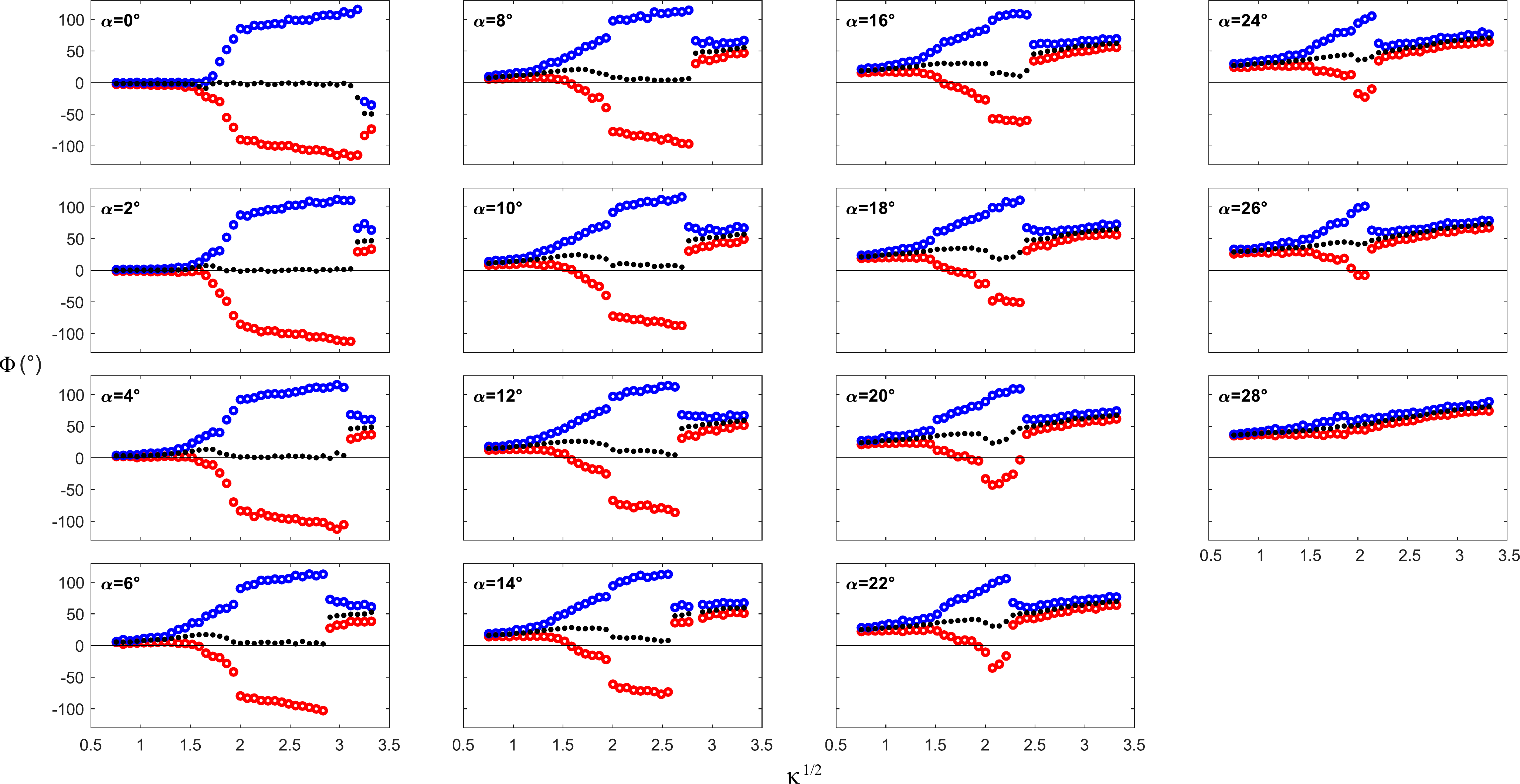}
        \caption{Maximum (\textcolor{blue}{$\circ$}), minimum (\textcolor{red}{$\circ$})  and mean ($\bullet$) deflection angle, $\Phi$, measured experimentally for an inverted cantilever plate of $\mathrm{AR}=5$ and $\mu=3.03$ as a function of non-dimensional flow speed, $\kappa$, and angle of attack, $\alpha$ (deg).}
        \label{phi5}
\end{figure}
\end{landscape}
\begin{landscape}
\begin{figure}
	\includegraphics[height=0.8\textwidth,keepaspectratio]{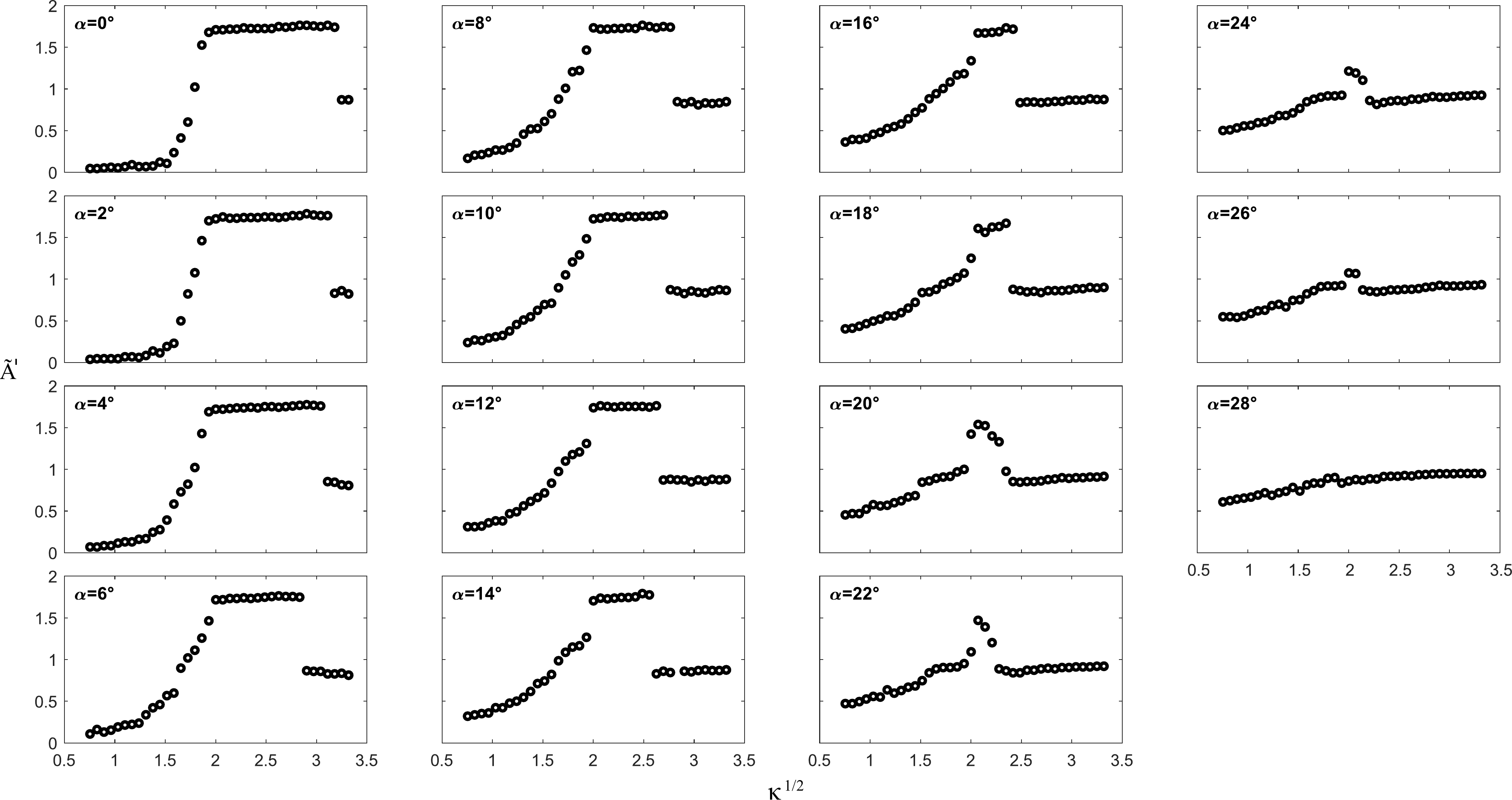}
        \caption{Non-dimensional maximum amplitude, $A'$, measured experimentally for an inverted cantilever plate of $\mathrm{AR}=5$ and $\mu=3.03$ as a function of non-dimensional flow speed, $\kappa$, and angle of attack, $\alpha$ (deg).}
        \label{A5}
\end{figure}
\end{landscape}
\begin{landscape}
\begin{figure}
	\includegraphics[height=0.8\textwidth,keepaspectratio]{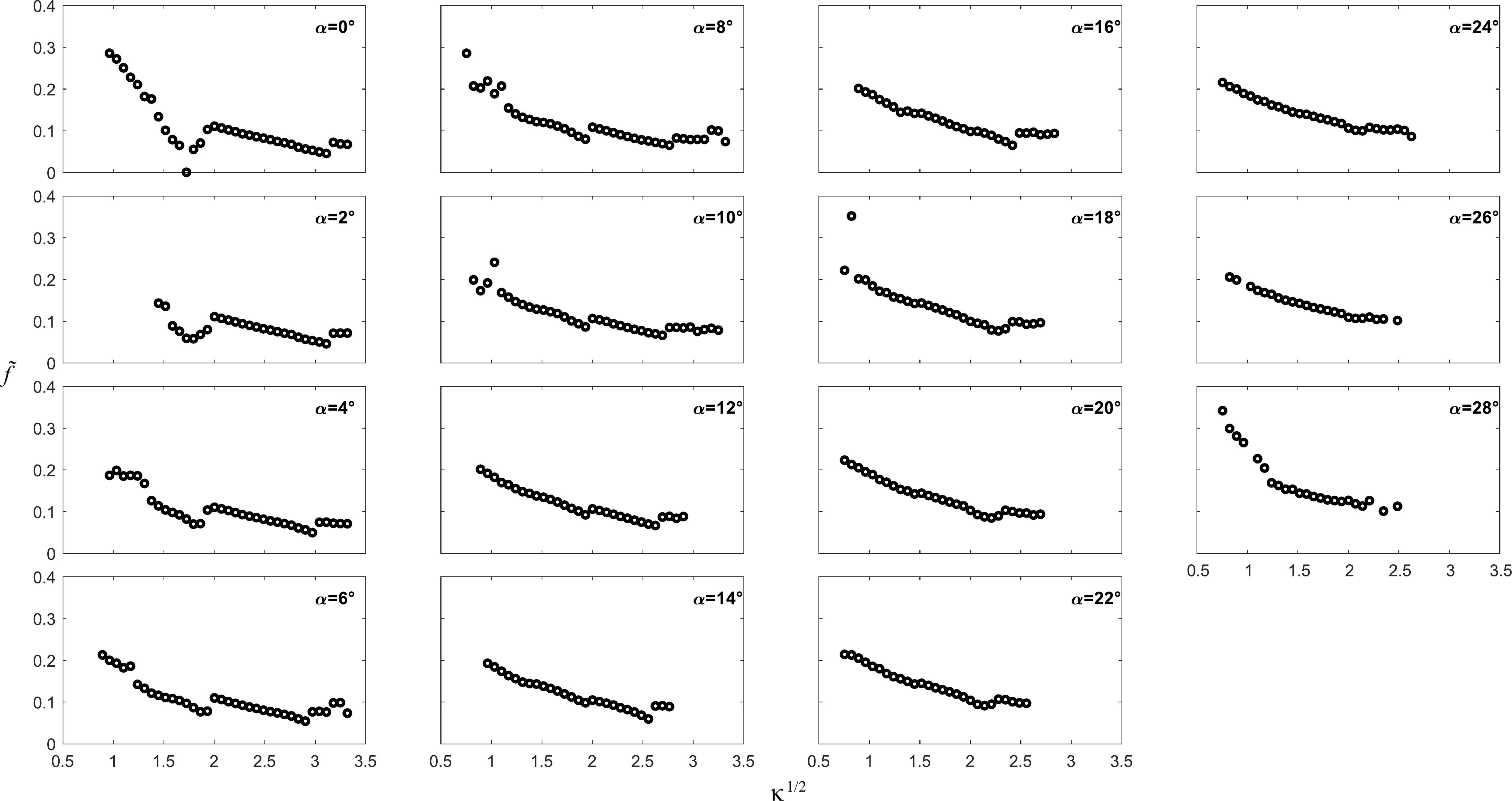}
        \caption{Non-dimensional frequency of motion, $\Tilde{f}$, measured experimentally for an inverted cantilever plate of $\mathrm{AR}=5$ and $\mu=3.03$ as a function of non-dimensional flow speed, $\kappa$, and angle of attack, $\alpha$ (deg).}
        \label{f5}
\end{figure}
\end{landscape}
\begin{landscape}
\begin{figure}
	\includegraphics[height=0.8\textwidth,keepaspectratio]{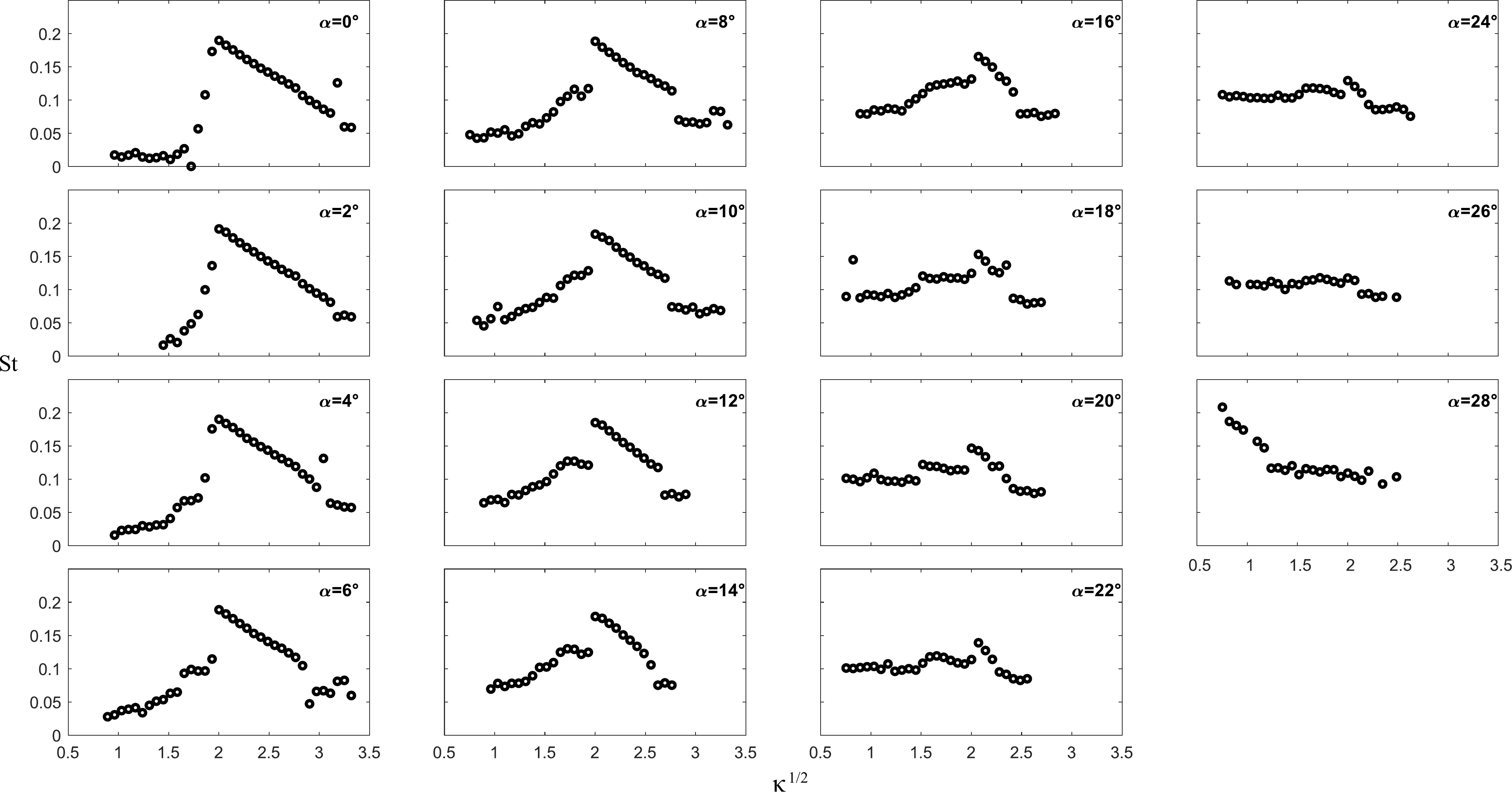}
        \caption{Strouhal number, $St=f A'/U$, measured experimentally for an inverted cantilever plate of $\mathrm{AR}=5$ and $\mu=3.03$ as a function of non-dimensional flow speed, $\kappa$, and angle of attack, $\alpha$(deg).}
        \label{St5}
\end{figure}
\end{landscape}

The dynamics of the $AR=5$ plate --- measured experimentally --- and those of the analogous 2D plate --- used in the numerical results --- will be analyzed at moderate angles of attack in the following sections. The case of $\alpha=\ang{10}$ will first be used as an example, because it embodies the general characteristics present for all angles of attack. The evolution of the plate's behavior as the angle of attack is increased will then be studied experimentally. The corresponding subsections will draw heavily from figures \ref{phi5}, \ref{A5}, \ref{f5} and \ref{St5}, which were obtained experimentally for $AR=5$ and angles of attack between \ang{0} and \ang{28}. Figure \ref{phi5} displays the maximum, minimum and average deflection angle as a function of non-dimensional flow speed. Each subfigure corresponds to a different angle of attack, in \ang{2} increments. The value of these angles is specified in the top left corner of each subfigure. Figure \ref{A5} follows a similar organization and shows the values of the amplitude $A'$, which, as specified in Section \ref{eq:parameters}, is obtained as the maximum between $A_{max}-A_{min}$ and $|A_{max}|$. Figure \ref{f5} shows, in a similar manner, the non-dimensional dominant frequency of the plate's motion, while figure \ref{St5} shows the Strouhal number $St= fA'/U$. As described above, the frequencies in these last two cases are obtained as the largest peak in the spectra of the flag's motion.  

\subsection{Deformed regime}

At small values of $\kappa^{1/2}$, inverted cantilever plates placed at zero angle of attack, $\alpha=\ang{0}$, exhibit a straight regime, characterized by small oscillations around the zero-deflection equilibrium. As is expected, inverted cantilever plates clamped at non-zero angles of attack do not exhibit this behavior. Instead, they undergo a small-amplitude oscillation around a small-deformation deflected equilibrium (figure \ref{modes}b), with the deflection of this equilibrium increasing as $\kappa^{1/2}$ is increased. This regime will be referred to as \textit{deformed} in this text, to make a distinction with the larger-deformation \textit{deflected} regime that arises at higher flow speeds above the lock-off of the flapping motion. Inverted cantilever plates clamped at $\alpha=\ang{0}$ transition from straight to deformed modes at a finite flow speed, while plates at a non-zero angles of attack are inherently in the deformed regime even at the smallest flow speeds.

The plate dynamics in the deformed regime are similar for all angles of attack, i.e., small oscillations around the deflected position (figure \ref{phi5}). The flow behavior, however, exhibits significant changes as the angle of attack and flow speed are increased. For small angles of attack ($\alpha \lesssim \ang{10}$) and low flow speeds, the deflection of the plate is small, and the flow remains attached. As flow speed is increased, the deflection increases accordingly and eventually reaches a critical value at which the flow separates. This transition towards separation and the subsequent initiation of vortex shedding marks the transition from the deformed regime to the small-amplitude flapping regime for plates at the low angles of attack ($\alpha \lesssim \ang{10}$), as will be shown in section \ref{AoA_evolution}. At high angles of attack ($\alpha \gtrsim \ang{12}$), on the other hand, the flow is separated for all flow speeds within the range considered in this study.

\begin{figure}
	\includegraphics[width=\textwidth]{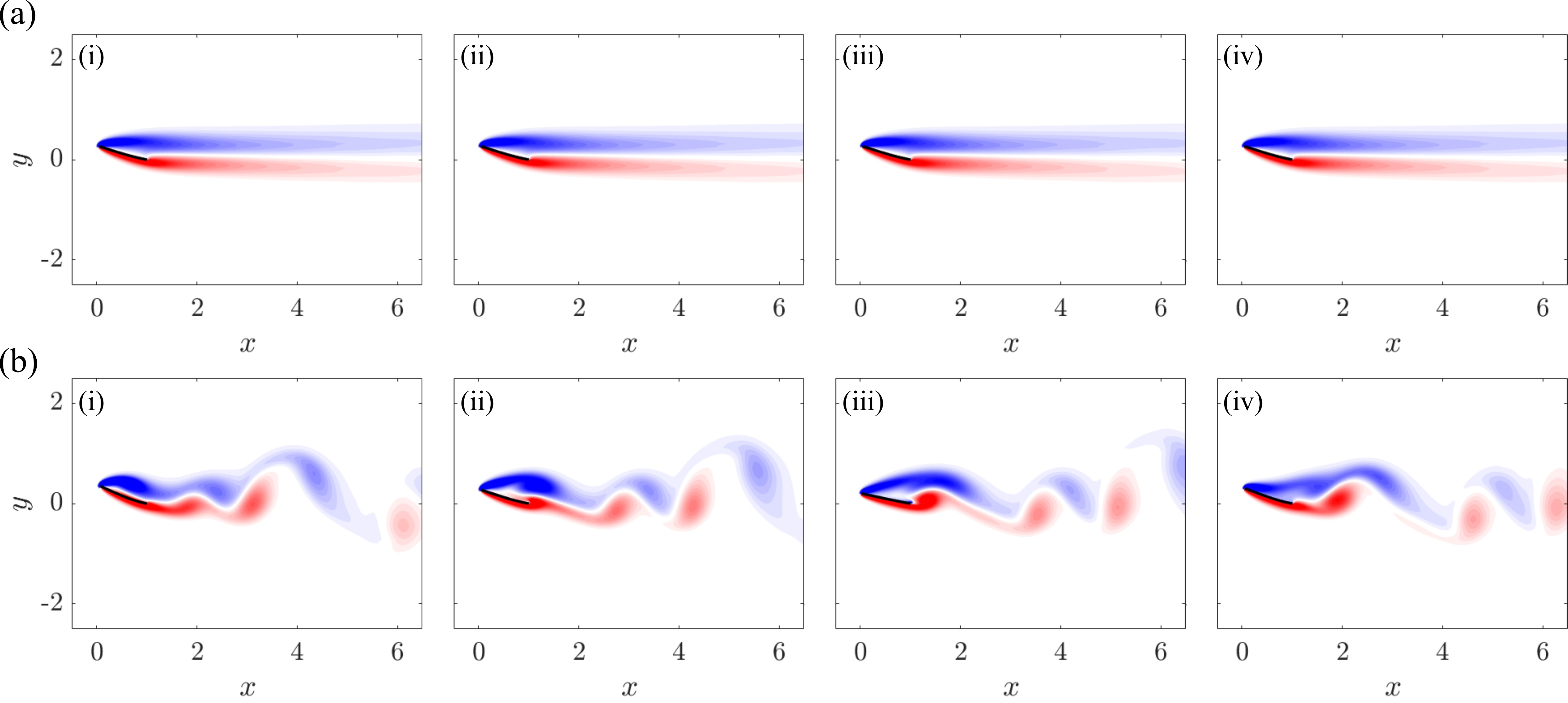}
        \caption[Snapshots]{Snapshots from the 2D numerical simulations for a plate at a) $\kappa^{1/2} = 1.15$  and b) $\kappa^{1/2} =1.25$. Snapshot i) corresponds to the plate being near its positive peak, and the remaining snapshots are spaced at roughly a quarter-period apart. Contours are of vorticity and are plotted in 34 levels between $-5$ and $5$ (these quantities are normalized using the freestream speed $U$ and plate length $L$).}
        \label{separation}
\end{figure}

To further demonstrate how flow separation demarcates the transition from deformed to small-amplitude regimes at small angles of attack, snapshots obtained from the numerical simulations at $\alpha=\ang{10}$ and $\kappa^{1/2}=1.15$ (in the deformed regime) and $\kappa^{1/2}=1.25$ (in the small-amplitude regime) are shown in figures \ref{separation}a and \ref{separation}b, respectively. The figures illustrate a full flapping cycle, with the first snapshot showing a flag position close to its positive peak and the remaining snapshots spaced roughly a quarter-period apart. For $\kappa^{1/2}=1.15$, long vortical structures extend into the wake, but no vortex shedding is evident. On the other hand, for $\kappa^{1/2}=1.25$ a 2S vortex shedding mode is clearly visible, despite the deflection of the plate being only slightly larger than for the $\kappa^{1/2}=1.15$ case. Notice by comparison with figure \ref{modes}e that this qualitative change in flow dynamics is commensurate with a jump in frequency of the flag's deflection dynamics. This jump in frequency is a result of the modification of aerodynamic forces when flow separation occurs. Thus, we utilize sharp jumps in frequency at low angles of attack, $\alpha$, and low flow speeds, $\kappa^{1/2}$, to indicate the transition from an attached to a detached flow. 

Using this criterion, the flow speed, $\kappa^{1/2}$, at which the flow separates for plates at small angles of attack can be identified in the experimental results in figure \ref{f5} for $\alpha=\ang{2}-\ang{10}$. The maximum deflection angles of the plate at the flow speeds for which this jump occurs in the experimental data are $\Phi_{sep}=\ang{8.8},\ \ang{9.8},\ \ang{13.0},\ \ang{12.4}$ and $\ang{15.1}$ at the angles of attack $\alpha=\ang{2},\ \ang{4},\ \ang{6},\ \ang{8}$ and $\ang{10}$, respectively. These deflection angles, $\Phi_{sep}$, are within the range expected for separation to occur. Variations in their value between the different angles of attack are a result of the varying plate geometry and flow speed at separation. By contrast, for the plates at larger angles of attack ($\alpha\gtrsim \ang{12}$), the flow is separated for all wind speeds within the studied range. Thus, even the deformed regime is marked by flow separation and vortex shedding processes. Consistent with this, there is no longer a jump in frequency in transitioning from the deformed to small-amplitude flapping regimes (figure \ref{f5}).

Finally, we note that the divergence instability that marks the onset of flapping for zero angle of attack, $\alpha=\ang{0}$ is no longer observed at non-zero angle of attack. As can be observed in figure \ref{f5}, the plate frequency does not approach $f=0$. Instead, as $\kappa$ is increased the amplitude of the plate oscillations begins to grow, and a large amplitude flapping motion develops. 

\subsection{Flapping regime} \label{IF_AoA_flapping}

\begin{figure}
	\includegraphics[width=\textwidth]{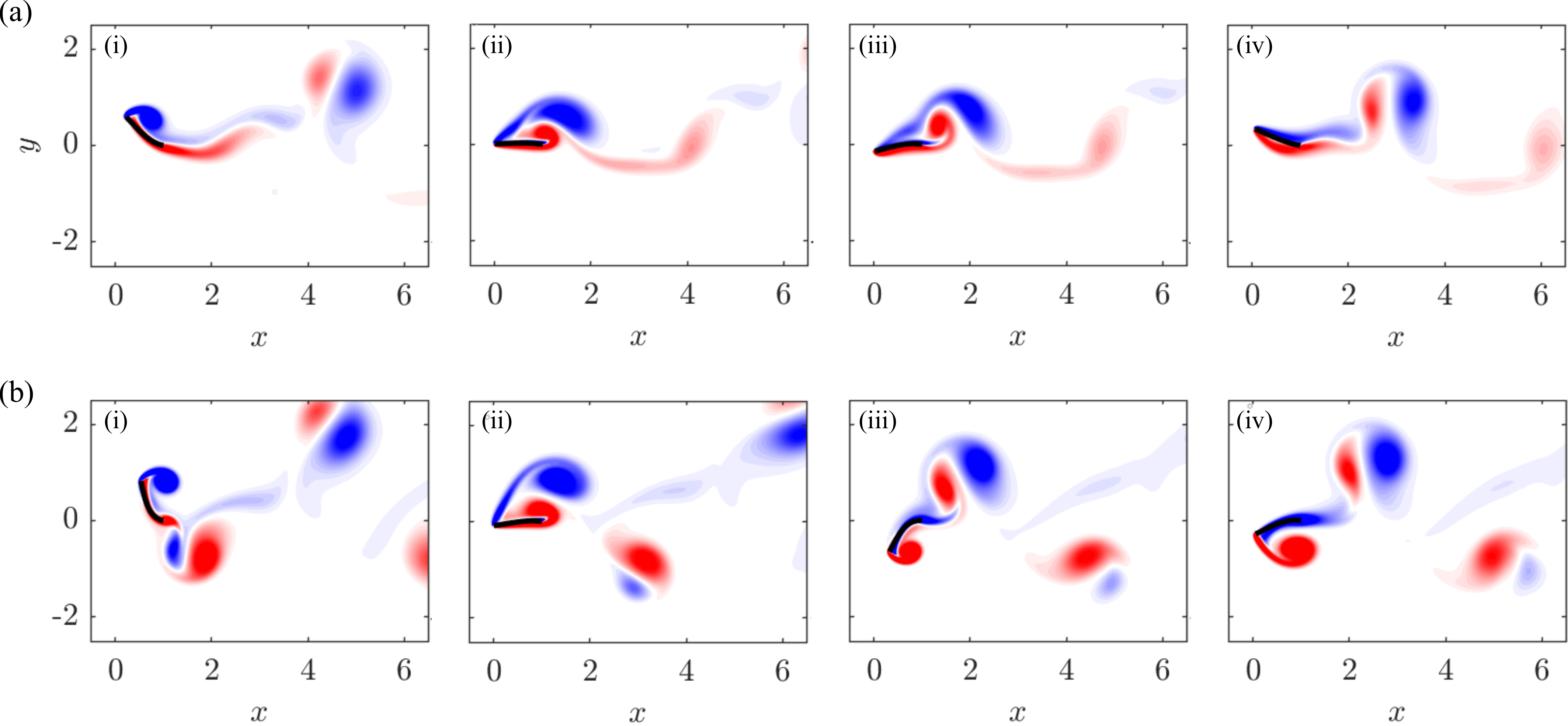}
        \caption[Snapshots]{Snapshots from the 2D numerical simulations for a plate at $\alpha=\ang{10}$ and a) $\kappa^{1/2} = 1.5$  and b) $\kappa^{1/2} =1.7$. Snapshot i) corresponds to the plate being near its positive peak, and the remaining snapshots are spaced at roughly a quarter-period apart. Contours are of vorticity and are plotted in 34 levels between $-5$ and $5$ (these quantities are normalized using the freestream speed $U$ and plate length $L$).}
        \label{snapshots}
\end{figure}

We denote the flapping regime as the collection of the small- and large-amplitude flapping regimes. The differences between these two regimes are easily recognizable in figure \ref{modes}. In the experimental results (figure \ref{modes}b), as example, the small-amplitude flapping regime is present between $\kappa_{lower}^{1/2}=1.17<\kappa^{1/2}<2$ and the large-amplitude flapping regime lies between $2<\kappa^{1/2}<\kappa_{upper}^{1/2}=2.69$, where $\kappa_{lower}$ and $\kappa_{upper}$ are used to denote the critical $\kappa$ for the initiation and cessation of flapping, respectively. The two regimes are separated by an abrupt change in amplitude, frequency and Strouhal number. Additionally, snapshots of the plate's motion in these flapping regimes, obtained through numerical simulations, are presented together with the corresponding vorticity contours in figure \ref{snapshots}. They illustrate a full flapping cycle, with the first snapshot showing a flag position close to its positive peak and the remaining snapshots spaced roughly a quarter-period apart. They are obtained after allowing sufficient time for the flag to reach periodic flapping motion. Figure \ref{snapshots}a corresponds to a flow speed in the small-amplitude flapping regime just before the transition to the large-amplitude flapping regime ($\kappa^{1/2}=1.5$), while figure \ref{snapshots}b corresponds to a flow speed just after the transition to large-amplitude flapping ($\kappa^{1/2}=1.7$). From these snapshots it is evident that, in addition to the amplitude, frequency and Strouhal number, the flow behavior is also disparate between small- and large-amplitude flapping regimes. 

The large-amplitude flapping regime is equivalent to the large-amplitude flapping motion present at zero angle of attack, which has been thoroughly described in the inverted-flag literature. It possesses all of the traits of a vortex-induced vibration, as established by \cite{Sader2016a}. The peak Strouhal number occurs at $\kappa^{1/2}=2$ in the experimental results, and is between St=0.18 -- 0.19 for angles $\alpha\lesssim\ang{15}$, decreasing rapidly for larger angles of attack (figure \ref{St5}). These Strouhal numbers are characteristic of lock-on in VIVs, marking the synchronization of vortex shedding with the resonant frequency of the plate. They decrease as flow speed is increased as a result of the commensurate decrease in frequency of oscillation (figure \ref{f5}), which is characteristic of vortex-induced vibrations in heavy fluid loading, and constant amplitude $A'$ (figure \ref{A5}), which is a product of the problem's geometry (increased deflections only result in the plate bending backwards beyond $\Phi=\ang{90}$). The large-amplitude flapping regime is associated with a 2P vortex shedding mode, as is evident in the numerical simulations of figure \ref{snapshots}b, in a similar manner to the large amplitude flapping mode of the zero angle of attack case \citep{Goza2018,Yu2017}. 

The small-amplitude flapping regime, on the other hand, presents different characteristics. The vortex dynamics are visible in the numerical simulations of figure \ref{snapshots}a. A single vortex pair is shed as the flag moves from the positive to the negative deflection peak. No vortex structures form, however, on the second half of the cycle, despite the absolute value of the minimum deflection being large ($\Phi=\ang{-21}$). While some vorticity is generated at the leading edge during this lower motion, no coherent vortex core is formed. The overall shedding pattern therefore corresponds to a 2S mode. This mode can also be observed at the lowest flow speed of the small-amplitude flapping regime, $\kappa^{1/2}=1.25$ in figure \ref{separation}b, and can, therefore, be expected to be present throughout the entirety of the regime. 

The small- and large-amplitude flapping motions of the inverted cantilever plate show striking similarities to the initial and lower branches present in the vortex-induced vibrations of a circular cylinder \citep{Williamson2004,Sarpkaya2004}. However, the $St$ within the small-amplitude flapping regime, determined experimentally using the frequency of motion, varies between St=0.02 and St=0.13 while the $St$ of a rigid flat plate at an angle of attack, calculated using the shedding frequency, is approximately $St \approx 0.15$ \citep{Knisely1990}. This significantly lower value of the frequency of motion compared to the vortex shedding frequency is not indicative of a classic vortex-induced vibration \citep{Williamson2004,Sarpkaya2004}. Although a more comprehensive analysis is required to fully discard a VIV as origin of the small amplitude flapping motion ---the difference in geometry between rigid and deformed plate may be behind this disparity in relative frequency values---the discrepancies point towards an origin of the small-amplitude flapping motion different from a VIV. The spectra of the plate's motion for flow speeds leading to the appearance of the small-amplitude flapping mode, visible in figure \ref{fft10} for the experimental results at $\alpha=\ang{10}$, show a single peak increasing in amplitude as the critical flow speed $\kappa_{lower}$ is approached, and not two coalescing peaks. Therefore, the presence of a coupled-mode instability, such as coupled-mode flutter, is unlikely. The most probable mechanism is that the plate's motion is dominated by the quasi-steady fluid mechanics, with the plate undergoing a limit-cycle oscillation similar to stall flutter. Further investigations, however, are necessary to unequivocally characterize the nature of this instability. It should be noted that plates at higher angles of attack ($\alpha>\ang{16}$) exhibit a discontinuity in flapping amplitude and oscillation frequency at a constant value of $\kappa^{1/2}\approx1.5$, which may be indicative of a transition to a different driving mechanism within the small-amplitude flapping regime.

\subsection{Deflected regime} \label{AoA_deflected}

In the large-amplitude flapping regime, the frequency of motion of the inverted cantilever plate decreases as flow speed is increased (figure \ref{f5}). When the value of the plate's resonant frequency becomes sufficiently different to that of the vortex shedding frequency, the plate's motion ceases to lock-on and the large-amplitude flapping motion disappears, giving rise to the deflected regime. This occurs at $St \simeq 0.08$ for the smallest angles of attack ($\alpha \leqslant \ang{4}$) and at $St \simeq 0.11$ for the larger angles (figure \ref{St5}). Between the flapping and deflected regimes, a bi-stable region, where either regime can exist, is experimentally observed. In this region, two different behaviors can occur. In the first case, the plate is either continuously flapping or continuously deflected depending on its initial condition. In the second case, the plate switches what appears to be randomly from one state to the other, resulting in the chaotic regime that has been reported for inverted flags at zero angle of attack \citep{Sader2016a,Goza2018}. This bi-stable region is observed to exist only for small angles of attack $\alpha\lesssim \ang{8}$ and occurs for a narrow band of flow speeds. Interestingly, the spectra of the plate's motion (figure \ref{fft10}) transitions from presenting a single dominant peak in the deformed and flapping regimes to presenting two clear peaks in the deflected regime. The frequency of the second peak, however, is too small to correspond to the vortex shedding frequency. It may correspond to either sub-harmonics of unsteady fluid forces or a resonant frequency of the plate.

The critical flow speed, $\kappa_{upper}$, at which the plate enters the deflected regime decreases with angle of attack (figure \ref{phi5}). \cite{Cosse2014b} observed that, independently of angle of attack, the inverted cantilever shows a similar deflected shape at the emergence of the deflected regime. This is corroborated in the present experimental measurements. The average deflection angle, $\bar{\Phi}$ at the smallest flow speed where the plate first enters the deflected regime is virtually constant, with an average of $\bar{\Phi}=\ang{46.7}$ and a standard deviation of only $\pm \ang{1.2}$. A similar value of $\bar{\Phi}=\ang{48.23}\pm\ang{3.42}$ is observed for the plate of $AR=2$ that will be discussed in section \ref{AR2}. The value of this angle for the numerical results obtained for $\alpha=\ang{10}$ is, on the other hand $\bar{\Phi}=\ang{61}$, with the difference in angle with respect to the experimental results being most likely caused by the differences in $Re$ and $\mu$. 

\section{Evolution with angle of attack} \label{AoA_evolution}

The four main dynamical regimes present at moderate angles of attack (deformed, small-amplitude flapping, large-amplitude flapping and deflected) have been analyzed in detail in the previous sections. The general evolution of the plate's dynamics with increasing angle of attack is visible in figures \ref{phi5}-\ref{St5}. The evolution of the critical transition speeds between the different regimes is analyzed in detail in this section. 

\begin{figure}
\centering
	\includegraphics[width=0.35\textwidth]{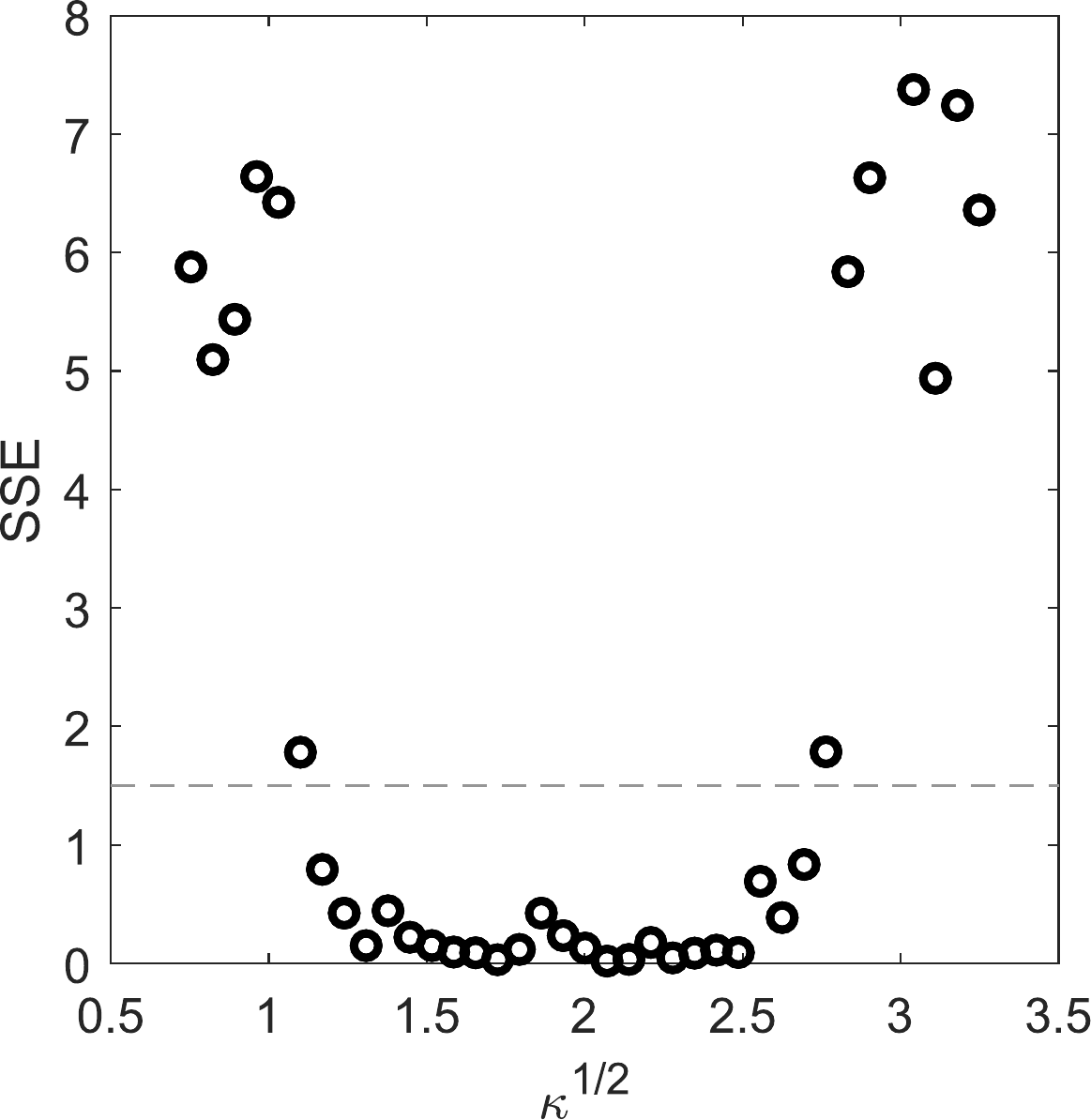}
        \caption{Sum of squares error for a bi-Gaussian fit to the power spectra of the plate's motion ($\circ$) and threshold defining the flapping regime (-) measured experimentally for an angle of $\alpha=\ang{10}$.}
        \label{SSE}
\end{figure}

The absence of a divergence instability for $\alpha>\ang{0}$ and the gradual increase of the plate's oscillation amplitude as flow speed is increased pose a challenge in defining a critical transition speed, $\kappa_{lower}$, from deformed to small-amplitude flapping regimes. The method proposed by \cite{Cosse2014} specifies $\kappa_{lower}$ as the speed at which the plate reaches an amplitude of motion that is a chosen fraction of the maximum flapping amplitude. The selection of this fraction, however, is arbitrary, and its variation results in significant changes in the value of the critical speed. An alternate approach, based on the spectra of the plate's deflection angle, is proposed here. The difference between flapping motions and motions with no resonance is distinguishable by eye in figure \ref{fft10}. In the flapping motions, the spectra show a single crisp peak that corresponds to the resonant frequency, while in the motions with no resonance the spectrum appears noisy due to the presence of multiple frequencies, even if a peak is present. The flapping regime is therefore defined as the range of flow speeds at which these spectra present a single crisp peak. In order to mathematically define this region, a bi-Gaussian function is fit to each FFT, normalized such that its maximum value is equal to one, and the sum of squares error (SSE) is calculated as

\begin{equation}
    SSE=\sum_1^{n}(y_i-\psi(f_i))^2
\end{equation}

\noindent where $n$ is the number of frequency points in the FFT, $y_i$ is the value of the FFT at the frequency $f_i$ and $\psi(f_i)$ is the value of the fit at $f_i$. The SSE provides a measure of the dispersion of the data points around the fit. The values of the SSE for $\alpha=\ang{10}$ are plotted in figure \ref{SSE}a. The SSE displays a low-value plateau at the flow speeds at which flapping occurs, with its value increasing in the deformed and deflected regimes. Consequently, the flapping regime is chosen to be the range of $\kappa$ at which the SSE is within this low-value plateau

\begin{equation}
\label{flapcondition}
SSE \left( \frac{FFT}{max(FFT)} \right) <1.5
\end{equation}

Although different angles of attack presented different defining thresholds, the value of 1.5 was chosen as a reasonable representative value for all angles. This threshold is considered less arbitrary than the maximum-amplitude fraction employed by \cite{Cosse2014} because it is chosen to separate the more clearly defined plateau. Alternative approaches to identifying this flapping region are certainly possible, and a more rigorous approach may be developed as our knowledge of the underlying physics expands. Even so, the principal conclusions presented in this section are expected to hold. The FFTs corresponding to the flapping regime, as defined by equation (\ref{flapcondition}), are identified with a bold frame in figure \ref{fft10}, for $\alpha=\ang{10}$. The corresponding critical flow speeds for the onset and end of the flapping regime are marked with vertical dashed lines in figure \ref{modes}b. While the transition from deformed to flapping regimes is smooth, the transition from flapping to deflected regimes is well defined and corresponds to an abrupt decrease in the amplitude of motion. The limits defined by equation (\ref{flapcondition}) capture this transition and the overall flapping region well.  

\begin{figure}
    \centering
	\includegraphics[width=0.6\textwidth]{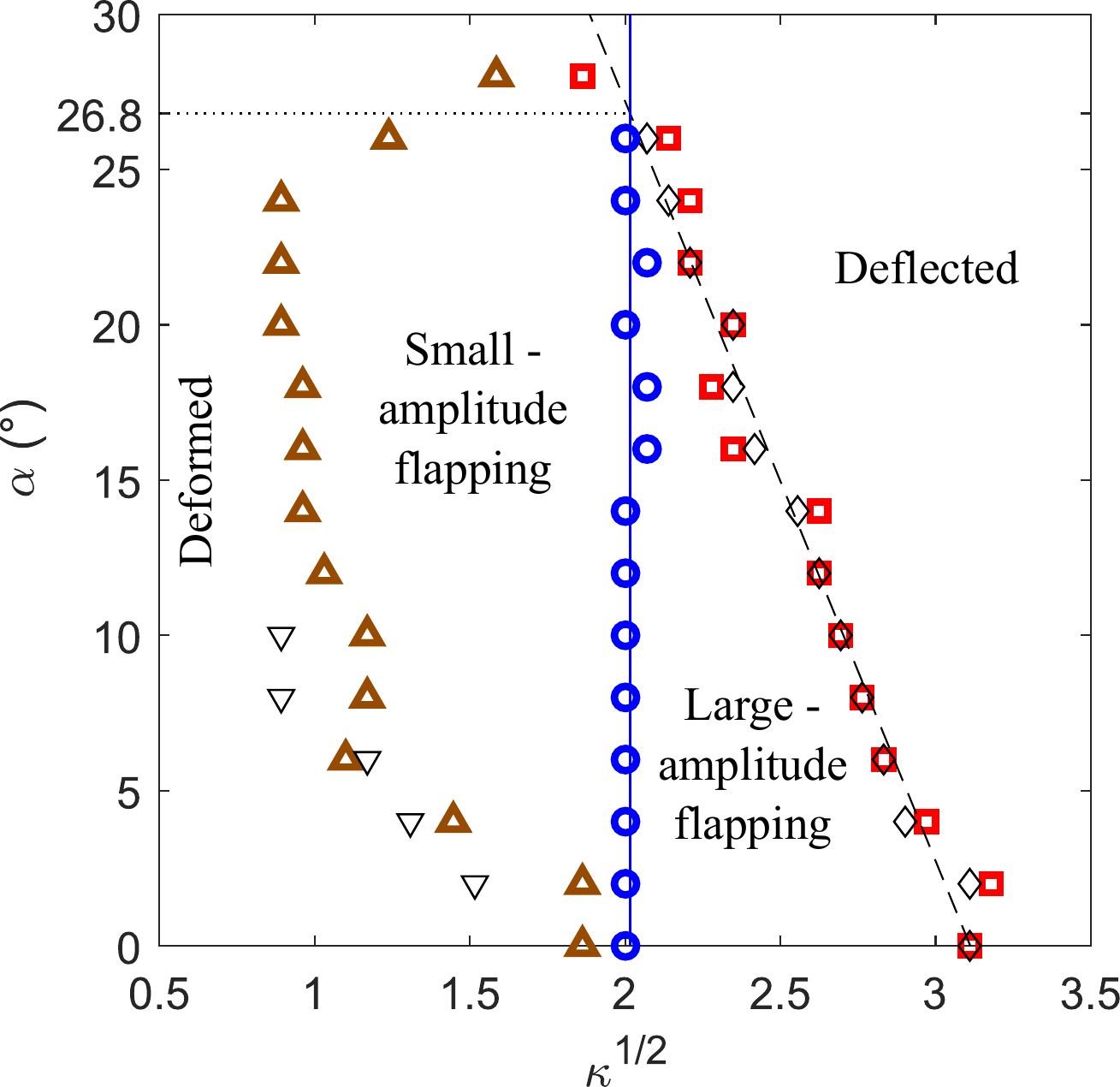}
        \caption{Critical non-dimensional flow speeds measured experimentally as a function of angle of attack. Flow speed for flow separation, $\kappa_{sep}$ ($\nabla$), start of small-amplitude flapping regime (as defined by equation (\ref{flapcondition})), $\kappa_{upper}$ (\textcolor{brown}{$\triangle$}), end of resonance (as defined by the same equation), $\kappa_{lower}$ (\textcolor{red}{$\square$}), start of deflected regime, $\kappa_{def}$ ($\diamond$) and start of large-amplitude flapping regime (\textcolor{blue}{$\circ$}). }
        \label{ksynch}
\end{figure}

The critical flow speeds at which transitions between regimes occur are represented in figure \ref{ksynch} for different angles of attack. The green triangles are $\kappa_{lower}$, which is the transition from deformed to small-amplitude flapping regimes as defined by equation \ref{flapcondition}. The red squares are $\kappa_{upper}$, the flow speed at which resonance ends, again using equation \ref{flapcondition}. The black triangles represent $\kappa_{sep}$, the flow speed at which the flow separates, which corresponds to a jump in frequency in figure \ref{f5} for angles $\alpha \leqslant \ang{10}$. The blue circles delineate the transition from small- to large-amplitude flapping, determined by a jump in amplitude and frequency in figures \ref{phi5}-\ref{St5}. The black rhombuses correspond to the flow speed at which large-amplitude flapping gives way to the deflected regime, $\kappa_{def}$, marked by a sharp decrease in amplitude in figure \ref{phi5}. 

The critical flow speed for transition from deformed to small-amplitude flapping regimes, $\kappa_{lower}$, shows different behavior for small ($\alpha \lesssim \ang{10}$) and large ($\alpha \gtrsim \ang{10}$) angles of attack. At small angles, it follows the trend of the flow speed required for the flow to separate, $\kappa_{sep}$, decreasing as angle of attack is increased (figure \ref{ksynch}). Following the discussion in section \ref{IF_AoA_flapping}, it is reasonable to assume a separated flow is required for the small-amplitude flapping regime to develop. This plot corroborates that at the emergence of small-amplitude flapping the flow over the plate is always separated. Thus, vortex shedding occurs and the quasi-steady forcing is non-linear. For larger angles of attack ($\alpha \gtrsim \ang{10}$), on the other hand, the flow is always separated within the range of $\kappa$ studied. At low $\kappa$ the plate remains in the deflected regime despite the flow being separated. Therefore, there must be other requirements in addition to flow separation for the small-amplitude flapping regime to develop, which may be related to the stability of the deformed position. The critical flow speed at these larger angles decreases significantly slower as the angle of attack is increased. At the highest angles of attack ($\alpha \gtrsim\ang{24}$), the transition flow speed increases abruptly and rapidly, leading to disappearance of the small-amplitude flapping regime.

The critical flow speed at which the plate enters the deflected regime, $\kappa_{def}$, and the critical flow speed at which the resonance ceases, $\kappa_{upper}$, are plotted together in figure \ref{ksynch}. Both speeds have been extracted from the same data set, eliminating any differences caused by the bi-stable nature of the plate in this region. As expected, these two speeds overlap, with small disparities being likely caused by experimental error. The critical flow speed at which the flag transitions to the deflected regime, $\kappa_{def}$, decreases as the angle of attack is increased, most probably due to the increased fluid damping. This decrease in $\kappa_{def}$ is linear with angle of attack. 

Moreover, the critical flow speed at which the plate enters the large-amplitude flapping regime, represented by blue circles in figure \ref{ksynch}, is strikingly constant for all $\alpha$, at a value of $\kappa^{1/2}=2$. Linear fits to both critical flow speeds have been included in figure \ref{ksynch}. Due to these trends, the range of flow speeds at which the large-amplitude flapping motion occurs decreases with increasing angle of attack. The large-amplitude flapping regime ceases to exist at an angle of $\alpha=\ang{26.8}$, calculated using the linear fits. The small-amplitude flapping motion is present at angles beyond this $\alpha=\ang{26.8}$ value, but not significantly higher. It ceases to exist for $\alpha\approx\ang{28}$. At larger angles ($\alpha\gtrsim\ang{28}$) the deformed and deflected regimes merge into a single common regime, where the plate flexes with continuously increasing deflection angle and oscillates with small amplitude around this position.

The angular amplitude of motion in the large-amplitude flapping regime is approximately independent of angle of attack for $\alpha\leqslant \ang{14}$. It decreases rapidly for angles greater than that value (figure \ref{phi5}). This result is in agreement with the threshold obtained by \cite{Shoele2016}, who observed the amplitude to notably decline for angles beyond $\alpha=\ang{15}$. The Strouhal number follows a similar trend, diminishing rapidly for angles greater than $\alpha=\ang{14}$ (figure \ref{St5}). The energy harvesting performance of the inverted flag is therefore severely limited beyond $\alpha=\ang{14}$, both due to the decrease in the flapping amplitude and the decrease in range of speeds at which flapping occurs.

\section{Effect of aspect ratio} \label{AR2}

\begin{landscape}
\begin{figure}
	\includegraphics[height=0.78\textwidth,keepaspectratio]{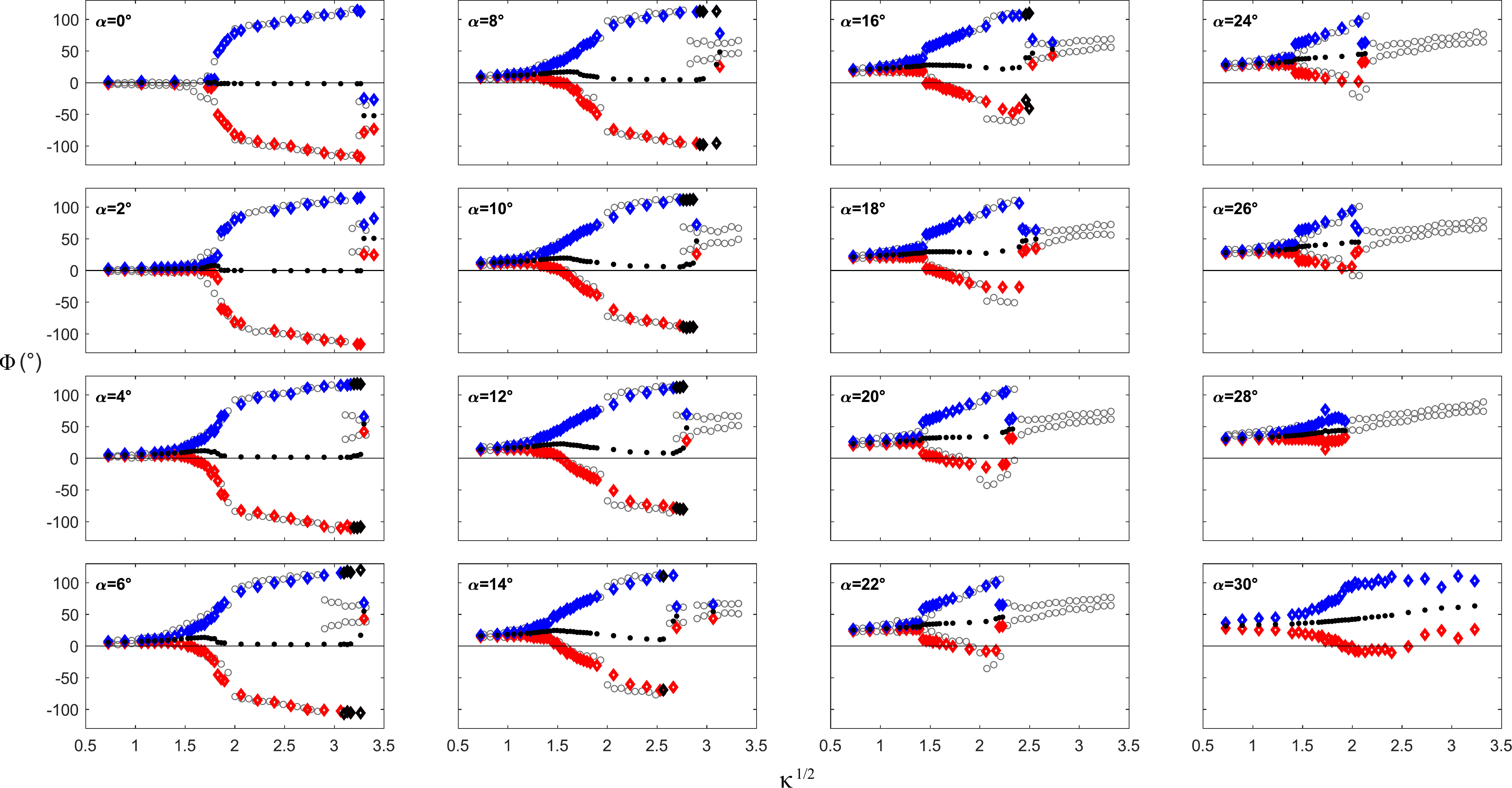}
        \caption{Maximum (\textcolor{blue}{$\diamond$}), minimum (\textcolor{red}{$\diamond$}) and mean ($\bullet$) deflection angle, $\Phi$, measured experimentally for an inverted cantilever plate of AR=2 and $\mu=3.11$ as a function of non-dimensional flow speed, $\kappa$, and angle of attack, $\alpha$. Maximum and minimum deflection angle for an inverted cantilever plate of AR=5 ($\circ$).}
        \label{phi16}
\end{figure}
\end{landscape}

\begin{landscape}
\begin{figure}
	\includegraphics[height=0.8\textwidth,keepaspectratio]{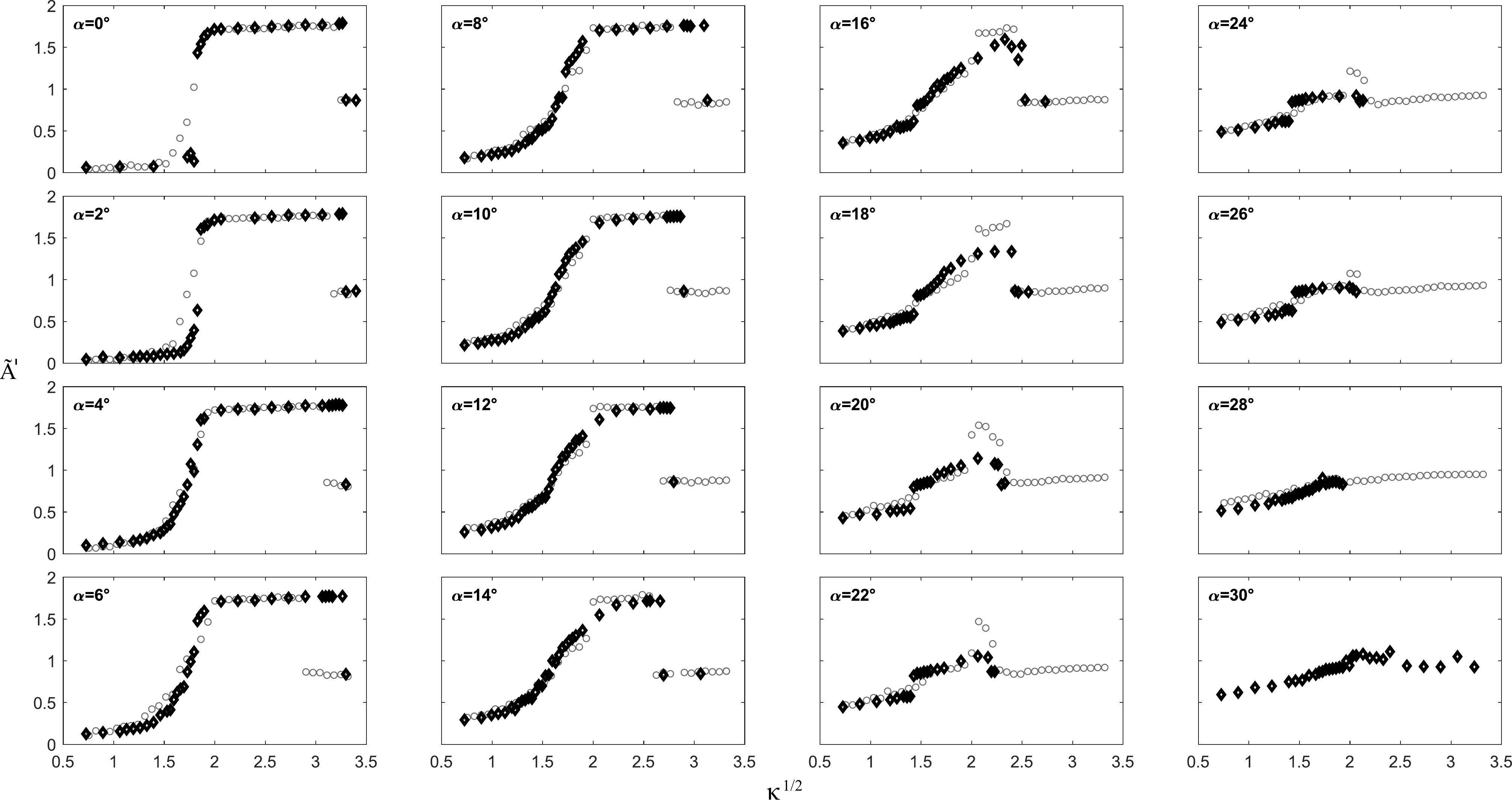}
        \caption{Maximum non-dimensional amplitude, A'($\diamond$), measured experimentally for an inverted cantilever plate of AR=2 and $\mu=3.11$ as a function of non-dimensional flow speed, $\kappa$, and angle of attack, $\alpha$. Maximum amplitude for an inverted cantilever plate of AR=5 ($\circ$) for reference. }
        \label{A16}
\end{figure}
\end{landscape}

\begin{landscape}
\begin{figure}
	\includegraphics[height=0.8\textwidth,keepaspectratio]{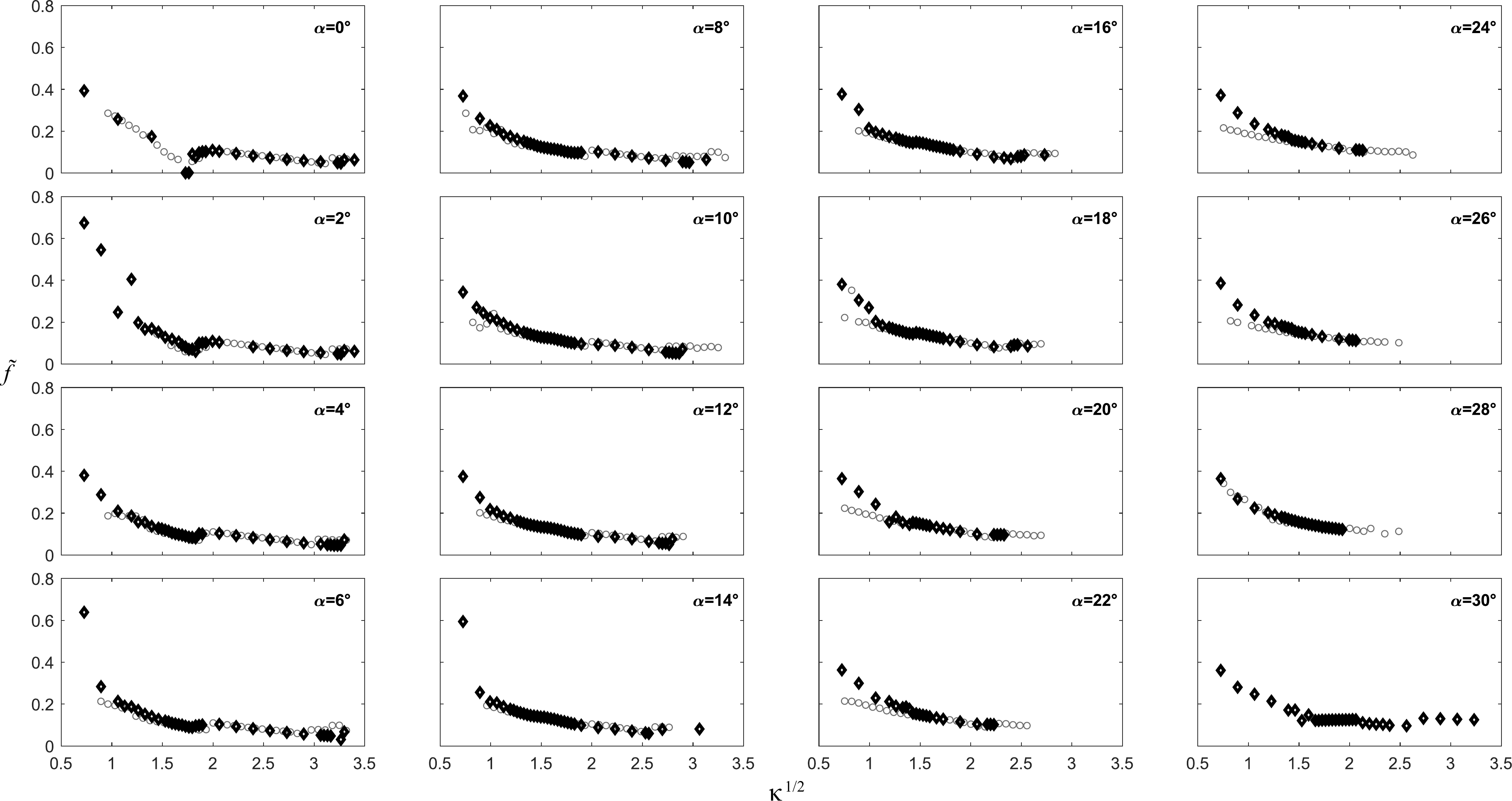}
        \caption{Non-dimensional frequency of motion, $\Tilde{f}$ ($\diamond$), measured experimentally for an inverted cantilever plate of AR=2 and $\mu=3.11$ as a function of non-dimensional flow speed, $\kappa$, and angle of attack, $\alpha$. Frequency of motion for an inverted cantilever plate of AR=5 ($\circ$) for reference.}
        \label{f16}
\end{figure}
\end{landscape}

\begin{landscape}
\begin{figure}
	\includegraphics[height=0.8\textwidth,keepaspectratio]{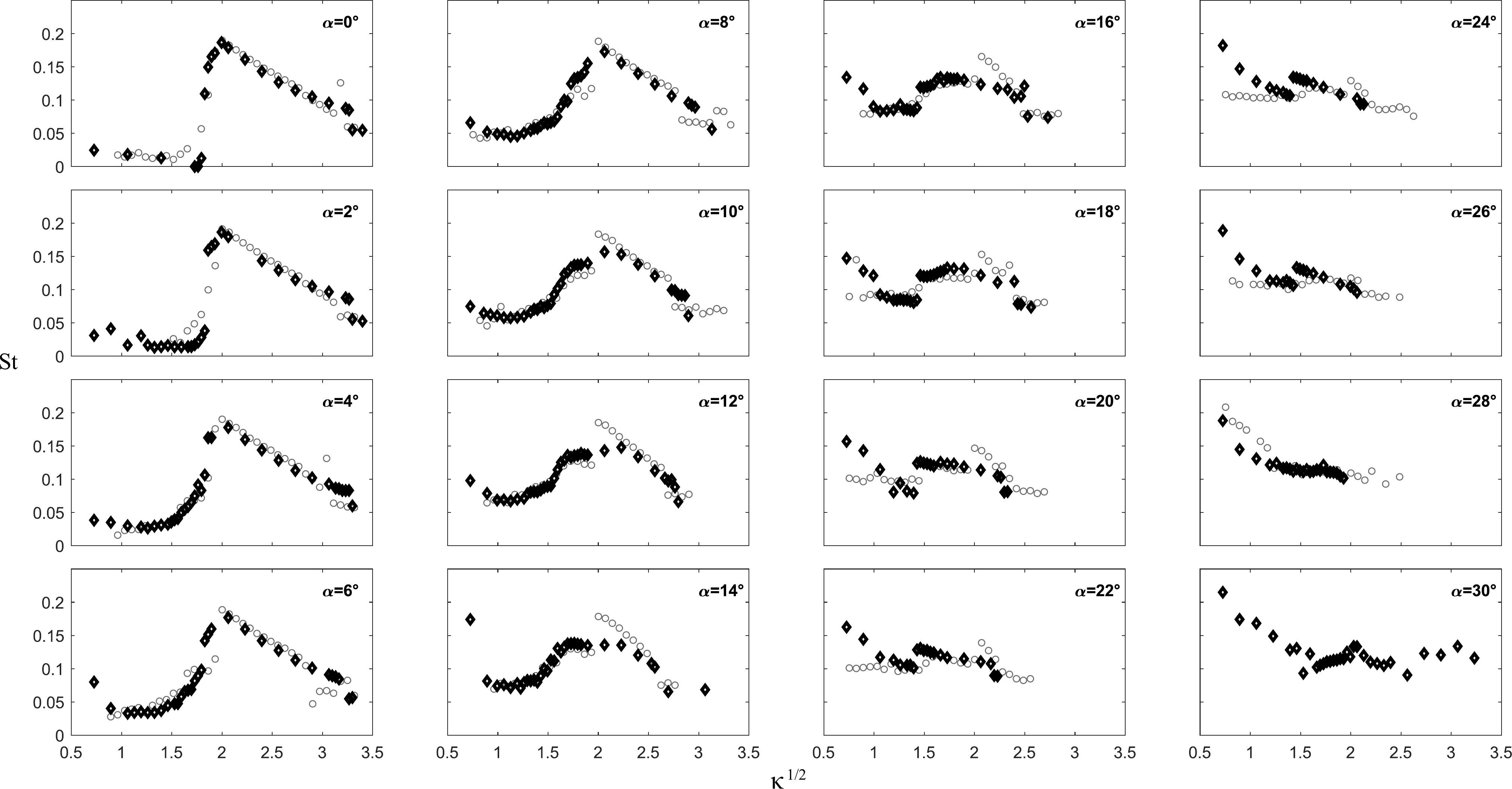}
        \caption{Strouhal number, $St=f A'/U$ ($\diamond$), measured experimentally for an inverted cantilever plate of AR=2 and $\mu=3.11$ as a function of non-dimensional flow speed, $\kappa$, and angle of attack, $\alpha$. Strouhal number for an inverted cantilever plate of AR=5 ($\circ$) for reference.}
        \label{St16}
\end{figure}
\end{landscape}

The experimental measurements reported in Section \ref{AR5} for an inverted cantilever plate of aspect ratio AR=5 are repeated here for a plate of AR=2, with the objective of highlighting the most prominent differences. The obtained data is presented in a similar manner: the maximum, minimum and mean deflection angle, $\Phi$, is shown in figure \ref{phi16}, the amplitude, $A'$, in figure \ref{A16}, the frequency, $\Tilde{f}$, in figure \ref{f16} and the Strouhal number, $St$, in figure \ref{St16}. The corresponding values for the AR=5 plate are included for comparison in these figures. The four main dynamical regimes present in the motion of the plate of AR=5 (deformed, small-amplitude flapping, large-amplitude flapping and deflected) can be recognized in the motion of the plate with AR=2. The presence of a chaotic regime is evident in the motion of this plate and occurs at a larger range of flow speeds than for AR=5. The data points corresponding to the chaotic region have been highlighted in black in figure \ref{phi16}.

\begin{figure}
	\includegraphics[width=\textwidth]{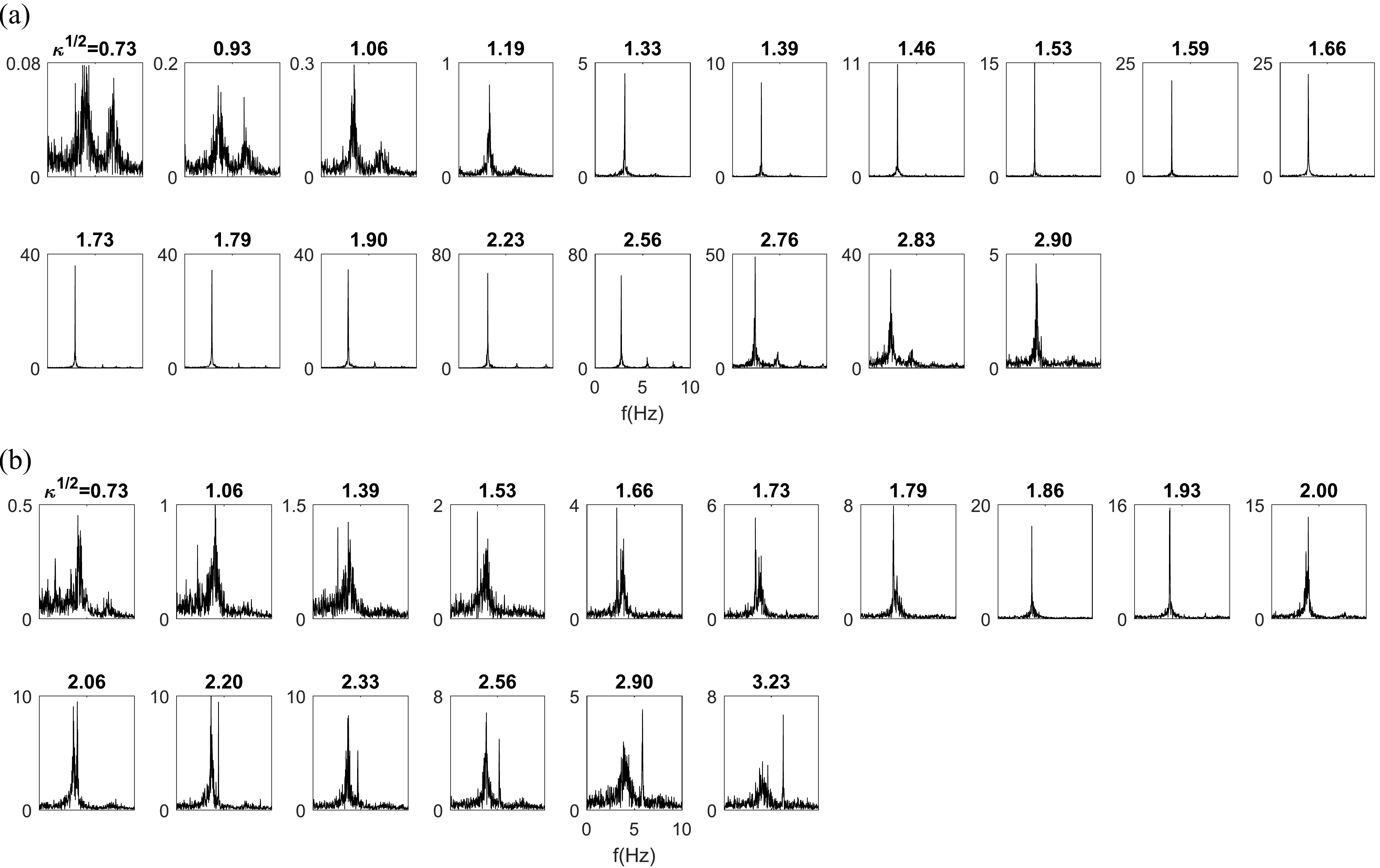}
        \caption{Power spectra of the deflection of an inverted cantilever plate of AR=2 at (a) $\alpha=\ang{10}$ and (b) $\alpha=\ang{30}$ obtained experimentally at varying flow speeds. The scale varies and is specified for each spectrum.}
        \label{fftAR2}
\end{figure}

At low flow speeds, the plate is in the deformed regime and oscillates with small amplitude around a small deflection equilibrium. In this regime the power spectra of the AR=2 plate, depicted in figure \ref{fftAR2}a, present two peaks, which may be indicative of the effect of the second length scale (height, H) that introduces a second characteristic frequency. The transition to the small-amplitude flapping regime occurs in a similar manner to the higher aspect ratio case, with the lower of the two frequency peaks growing in amplitude and without a coalescence of peaks being observed (figure \ref{fftAR2}a). The jump in frequency indicative of flow detachment at small angles of attack is, however, no longer evident.

The behavior of the plate in the small- and large-amplitude flapping regimes presents notable differences. For the plate of AR=2, an abrupt increase in amplitude and Strouhal number, signaling the transition from small- to large-amplitude flapping, is present only at the lower angles of attack ($\alpha\leqslant \ang{6}$). For these small angles, the transition does not occur at the flow speed where $St$ is maximum (figures \ref{St16}), which was the case for the plate of AR=5. The flow speed where the $St$ peaks is, however, strikingly equal for both aspect ratios ($\kappa^{1/2}\approx2$). For larger angles of attack ($\alpha > \ang{6}$), the plate of AR=2 does not present a clear transition from small to large amplitude flapping. The variables that define its motion (figures \ref{phi16} -- \ref{St16}) vary smoothly throughout the flapping regime. This may be caused by the existence of tip vortices and three-dimensionality in vortex shedding, that result in the presence of a range of frequency values instead of a single defined frequency. The absence of a pronounced large-amplitude flapping regime for the AR=2 case is marked by lower flapping amplitudes at angles of attack $\alpha>\ang{8}$, with the difference in minimum deflection angle being particularly significant (figure \ref{phi16}). The discrepancy in flapping amplitude between AR=2 and AR=5 initially occurs only at the lower flow speeds of the large-amplitude flapping regime, but extends to the full range of speeds for $\alpha>\ang{14}$. The significantly lowered amplitude for $\alpha>\ang{8}$ produces lower Strouhal numbers, making the nature of the flapping motion unclear. Remarkably, the plate of AR=5 presented an additional discontinuity in amplitude for angles $\alpha>\ang{16}$ at a flow speed of $\kappa^{1/2}\approx1.5$ that is also present at the same threshold and flow speed for AR=2. 

The flow speed at which the large-amplitude flapping regime ends for AR=2 follows a linear trend with angle of attack, but has a more pronounced slope than the AR=5 case. Similarly to the plate of AR=5, the flapping motion is practically non-existent for an angles of attack beyond $\alpha=\ang{28}$ for AR=2. An angle of attack of $\alpha=\ang{30}$ was additionally investigated, the data from which are presented in figures \ref{phi16}--\ref{St16}. Interestingly, a new resonant motion is observed that presents amplitude and frequency response characteristics distinct from those of the flapping regimes described above. At the lower flow speeds, the power spectra of the motion, displayed in figure \ref{fftAR2}b, show the existence of two peaks, which approach each other at the onset of the large-amplitude motion of the plate. The proximity of these peaks suggests the presence of a coupled-mode mechanism, such as coupled-mode flutter. The detailed analysis of this new resonant motion is beyond the present scope, but these results highlight the variety of phenomena that may arise for certain plate dimensions at angles of attack intermediate to those typically studied, $\alpha=\ang{0},\ \ang{90}$ and $\ang{180}$, that may be of interest in future investigations.

\section{Conclusions} \label{conclusions}

This study has reported and analyzed the dynamics of an inverted cantilevered plate placed at a moderate angle of attack ($\alpha \leqslant \ang{30}$) to a uniform flow. First, a plate of large aspect ratio (AR=5) was investigated experimentally. At non-zero angle of attack, four distinct dynamical regimes were identified as the non-dimensional flow speed, $\kappa$, is increased: deformed (small oscillations around a small deflection equilibrium), small-amplitude flapping, large-amplitude flapping and deflected (small oscillations around a large deflection equilibrium). A comparison with 2D numerical simulations revealed that the qualitative dynamical features remain unchanged for low Reynolds numbers ($Re=200$), despite variations in characteristic flow speeds, amplitudes and frequencies.

The data obtained for the AR=5 case showed that the plate, initially in the deformed regime, does not undergo a divergence instability, but exhibits a gradual increase in the amplitude of oscillation as it enters the small-amplitude flapping regime. A new method based on the power spectra of the plate's deflection was proposed to identify the critical flow speed at which small-amplitude flapping is initiated. For small angles of attack ($\alpha \leqslant \ang{10}$) this transition occurs at the flow speed that generates a plate deflection large enough for the flow to separate. For large angles of attack ($\alpha>\ang{10}$) the flow is always separated and an additional instability, which remains to be fully characterized, determines this critical flow speed. The small-amplitude flapping regime was argued to be a limit cycle oscillation produced by the quasi-steady fluid forcing and was associated to a 2S vortex shedding mode. 

At a flow speed of $\kappa^{1/2}=2$, which was found to be independent of angle of attack, the plate transitions to the large-amplitude flapping regime. This transition is marked by an abrupt change in flapping amplitude, frequency and Strouhal number as well as a shift to a 2P vortex shedding mode. The large-amplitude flapping regime presents the same characteristics as the flapping motion observed at zero angle of attack, which has been established to constitute a vortex-induced vibration \citep{Sader2016a}. As the flow speed was increased, the plate transitioned to the deflected regime. The flow speed at which this transition occurs decreases linearly with angle of attack. Consequently, the range of flow speeds at which the large-amplitude flapping motion is present decreases with increasing angle of attack. At an angle of attack of $\alpha=\ang{26.8}$ the large amplitude flapping motion ceases to exist. The small-amplitude flapping motion disappears at angles beyond $\alpha\approx\ang{28}$.

A subsequent set of measurements were performed on a plate of aspect ratio AR=2 and revealed contrasting dynamics. Although the main dynamic regimes are still present, the distinction between small and large amplitude flapping motions is no longer evident, with the amplitude and frequency of motion varying smoothly across the flapping regimes. This was argued to be a result of the three-dimensional vortex shedding. As a result of the modified dynamics, the flapping amplitude is fairly decreased with respect to the large aspect ratio plate for angles of attack above $\alpha\gtrsim\ang{8}$. A new distinct large-amplitude motion was observed to appear at $\alpha=\ang{30}$, highlighting the diversity of dynamics that may be present at larger angles of attack. 

\section{Acknowledgements}

C.H.-C. and M.G. acknowledge funding from the Gordon and Betty Moore Foundation. A.G. and T.C. acknowledge funding from Robert Bosch LLC through the Bosch Energy Research Network Grant (grant number 07.23.CS.15), and from the AFOSR (grant number FA9550-14-1-0328). J.S. acknowledges funding from the  Australian Research Council Centre of Excellence in Exciton Science (CE170100026) and the Australian Research Council grants scheme.

Declaration of Interests. The authors report no conflict of interest.

\bibliographystyle{jfm}

\bibliography{AoA_JFM_v11}

\begin{thebibliography}{39}
\expandafter\ifx\csname natexlab\endcsname\relax\def\natexlab#1{#1}\fi
\def\au#1{#1} \def\ed#1{#1} \def\yr#1{#1}\def\at#1{#1}\def\jt#1{\textit{#1}}
  \def\bt#1{#1}\def\bvol#1{\textbf{#1}} \def\vol#1{#1} \def\pg#1{#1}
  \def\publ#1{#1}\def\arxiv#1{#1}\def\org#1{#1}\def\st#1{\textit{#1}}

\bibitem[Alben \& Shelley(2008)]{Alben2008}
{\sc \au{Alben, Silas} \& \au{Shelley, Michael~J}} \yr{2008}  \at{Flapping
  states of a flag in an inviscid fluid: bistability and the transition to
  chaos}.  \jt{Physical review letters}  \bvol{100}~(7),  \pg{074301}.

\bibitem[Allen \& Smits(2001)]{Allen2001}
{\sc \au{Allen, JJ} \& \au{Smits, AJ}} \yr{2001}  \at{Energy harvesting eel}.
  \jt{Journal of fluids and structures}  \bvol{15}~(3-4),  \pg{629--640}.

\bibitem[Colonius \& Taira(2008)]{Colonius2008}
{\sc \au{Colonius, Tim} \& \au{Taira, Kunihiko}} \yr{2008}  \at{A fast immersed
  boundary method using a nullspace approach and multi-domain far-field
  boundary conditions}.  \jt{Computer Methods in Applied Mechanics and
  Engineering}  \bvol{197}~(25),  \pg{2131--2146}.

\bibitem[Coss{\'e} {\em et~al.\/}(2014)Coss{\'e}, Sader, Kim, Huertas~Cerdeira
  \& Gharib]{Cosse2014}
{\sc \au{Coss{\'e}, Julia}, \au{Sader, John}, \au{Kim, Daegyoum},
  \au{Huertas~Cerdeira, Cecilia} \& \au{Gharib, Morteza}} \yr{2014} The effect
  of aspect ratio and angle of attack on the transition regions of the inverted
  flag instability.  \bt{In {\em Proc. ASME\/}}.

\bibitem[Coss{\'e}(2014)]{Cosse2014b}
{\sc \au{Coss{\'e}, Julia~Theresa}} \yr{2014}  \at{On the behavior of pliable
  plate dynamics in wind: application to vertical axis wind turbines}. PhD
  thesis, California Institute of Technology.

\bibitem[Criesfield(1991)]{Crisfield1991}
{\sc \au{Criesfield, MA}} \yr{1991} {\em Non-linear finite element analysis of
  solids and structures, vol. 1\/}.  \publ{Wiley, New York}.

\bibitem[De~Langre(2008)]{Delangre2008}
{\sc \au{De~Langre, Emmanuel}} \yr{2008}  \at{Effects of wind on plants}.
  \jt{Annu. Rev. Fluid Mech.}  \bvol{40},  \pg{141--168}.

\bibitem[Driessen {\em et~al.\/}(2007)Driessen, Mol, Bouten \&
  Baaijens]{Driessen2007}
{\sc \au{Driessen, Niels~JB}, \au{Mol, Anita}, \au{Bouten, Carlijn~VC} \&
  \au{Baaijens, Frank~PT}} \yr{2007}  \at{Modeling the mechanics of
  tissue-engineered human heart valve leaflets}.  \jt{Journal of biomechanics}
  \bvol{40}~(2),  \pg{325--334}.

\bibitem[Eloy {\em et~al.\/}(2008)Eloy, Lagrange, Souilliez \&
  Schouveiler]{Eloy2008}
{\sc \au{Eloy, Christophe}, \au{Lagrange, Romain}, \au{Souilliez, Claire} \&
  \au{Schouveiler, Lionel}} \yr{2008}  \at{Aeroelastic instability of
  cantilevered flexible plates in uniform flow}.  \jt{Journal of Fluid
  Mechanics}  \bvol{611},  \pg{97--106}.

\bibitem[Fan {\em et~al.\/}(2019)Fan, Huertas-Cerdeira, Coss{\'e}, Sader \&
  Gharib]{Fan2019}
{\sc \au{Fan, Boyu}, \au{Huertas-Cerdeira, Cecilia}, \au{Coss{\'e}, Julia},
  \au{Sader, John~E} \& \au{Gharib, Morteza}} \yr{2019}  \at{Effect of
  morphology on the large-amplitude flapping dynamics of an inverted flag in a
  uniform flow}.  \jt{Journal of Fluid Mechanics}  \bvol{874},  \pg{526--547}.

\bibitem[Goza \& Colonius(2017)]{Goza2017}
{\sc \au{Goza, Andres} \& \au{Colonius, Tim}} \yr{2017}  \at{A strongly-coupled
  immersed-boundary formulation for thin elastic structures}.  \jt{Journal of
  Computational Physics}  \bvol{336},  \pg{401--411}.

\bibitem[Goza {\em et~al.\/}(2018)Goza, Colonius \& Sader]{Goza2018}
{\sc \au{Goza, Andres}, \au{Colonius, Tim} \& \au{Sader, John~E}} \yr{2018}
  \at{Global modes and nonlinear analysis of inverted-flag flapping}.
  \jt{Journal of Fluid Mechanics}  \bvol{857},  \pg{312--344}.

\bibitem[Goza {\em et~al.\/}(2016)Goza, Liska, Morley \& Colonius]{Goza2016}
{\sc \au{Goza, Andres}, \au{Liska, Sebastian}, \au{Morley, Benjamin} \&
  \au{Colonius, Tim}} \yr{2016}  \at{Accurate computation of surface stresses
  and forces with immersed boundary methods}.  \jt{Journal of Computational
  Physics}  \bvol{321},  \pg{860--873}.

\bibitem[Gurugubelli \& Jaiman(2015)]{Gurugubelli2015}
{\sc \au{Gurugubelli, PS} \& \au{Jaiman, RK}} \yr{2015}  \at{Self-induced
  flapping dynamics of a flexible inverted foil in a uniform flow}.
  \jt{Journal of Fluid Mechanics}  \bvol{781},  \pg{657--694}.

\bibitem[Huertas-Cerdeira {\em et~al.\/}(2018)Huertas-Cerdeira, Fan \&
  Gharib]{Huertas2018}
{\sc \au{Huertas-Cerdeira, Cecilia}, \au{Fan, Boyu} \& \au{Gharib, Morteza}}
  \yr{2018}  \at{Coupled motion of two side-by-side inverted flags}.
  \jt{Journal of Fluids and Structures}  \bvol{76},  \pg{527--535}.

\bibitem[Kim {\em et~al.\/}(2013)Kim, Coss{\'e}, Cerdeira \& Gharib]{Kim2013}
{\sc \au{Kim, Daegyoum}, \au{Coss{\'e}, Julia}, \au{Cerdeira, Cecilia~Huertas}
  \& \au{Gharib, Morteza}} \yr{2013}  \at{Flapping dynamics of an inverted
  flag}.  \jt{Journal of Fluid Mechanics}  \bvol{736},  \pg{R1}.

\bibitem[Knisely(1990)]{Knisely1990}
{\sc \au{Knisely, Charles~W}} \yr{1990}  \at{Strouhal numbers of rectangular
  cylinders at incidence: a review and new data}.  \jt{Journal of Fluids and
  Structures}  \bvol{4}~(4),  \pg{371--393}.

\bibitem[Lee {\em et~al.\/}(2015)Lee, Sherrit, Tosi, Walkemeyer \&
  Colonius]{Lee2015}
{\sc \au{Lee, Hyeong}, \au{Sherrit, Stewart}, \au{Tosi, Luis}, \au{Walkemeyer,
  Phillip} \& \au{Colonius, Tim}} \yr{2015}  \at{Piezoelectric energy
  harvesting in internal fluid flow}.  \jt{Sensors}  \bvol{15}~(10),
  \pg{26039--26062}.

\bibitem[Luhar \& Nepf(2011)]{Luhar2011}
{\sc \au{Luhar, M.} \& \au{Nepf, H.~M.}} \yr{2011}  \at{Flow-induced
  reconfiguration of buoyant and flexible aquatic vegetation}.  \jt{Limnol.
  Oceanogr.}  \bvol{56},  \pg{2003--2017}.

\bibitem[Orrego {\em et~al.\/}(2017)Orrego, Shoele, Ruas, Doran, Caggiano,
  Mittal \& Kang]{Orrego2017}
{\sc \au{Orrego, Santiago}, \au{Shoele, Kourosh}, \au{Ruas, Andre}, \au{Doran,
  Kyle}, \au{Caggiano, Brett}, \au{Mittal, Rajat} \& \au{Kang, Sung~Hoon}}
  \yr{2017}  \at{Harvesting ambient wind energy with an inverted piezoelectric
  flag}.  \jt{Applied Energy}  \bvol{194},  \pg{212--222}.

\bibitem[Paidoussis(1998)]{Paidoussis1998}
{\sc \au{Paidoussis, Michael~P}} \yr{1998} {\em Fluid-structure interactions:
  slender structures and axial flow\/}, ,  \vol{vol.~1}.  \publ{Academic
  press}.

\bibitem[Park {\em et~al.\/}(2016)Park, Kim, Chang, Ryu \& Sung]{Park2016}
{\sc \au{Park, Sung~Goon}, \au{Kim, Boyoung}, \au{Chang, Cheong~Bong}, \au{Ryu,
  Jaeha} \& \au{Sung, Hyung~Jin}} \yr{2016}  \at{Enhancement of heat transfer
  by a self-oscillating inverted flag in a poiseuille channel flow}.
  \jt{International Journal of Heat and Mass Transfer}  \bvol{96},
  \pg{362--370}.

\bibitem[Ryu {\em et~al.\/}(2015)Ryu, Park, Kim \& Sung]{Ryu2015}
{\sc \au{Ryu, Jaeha}, \au{Park, Sung~Goon}, \au{Kim, Boyoung} \& \au{Sung,
  Hyung~Jin}} \yr{2015}  \at{Flapping dynamics of an inverted flag in a uniform
  flow}.  \jt{Journal of Fluids and Structures}  \bvol{57},  \pg{159--169}.

\bibitem[Sader {\em et~al.\/}(2016{\natexlab{{\em a\/}}})Sader, Coss{\'e}, Kim,
  Fan \& Gharib]{Sader2016a}
{\sc \au{Sader, John~E}, \au{Coss{\'e}, Julia}, \au{Kim, Daegyoum}, \au{Fan,
  Boyu} \& \au{Gharib, Morteza}} \yr{2016{\natexlab{{\em a\/}}}}
  \at{Large-amplitude flapping of an inverted flag in a uniform steady flow--a
  vortex-induced vibration}.  \jt{Journal of Fluid Mechanics}  \bvol{793},
  \pg{524--555}.

\bibitem[Sader {\em et~al.\/}(2016{\natexlab{{\em b\/}}})Sader,
  Huertas-Cerdeira \& Gharib]{Sader2016b}
{\sc \au{Sader, John~E}, \au{Huertas-Cerdeira, Cecilia} \& \au{Gharib,
  Morteza}} \yr{2016{\natexlab{{\em b\/}}}}  \at{Stability of slender inverted
  flags and rods in uniform steady flow}.  \jt{Journal of Fluid Mechanics}
  \bvol{809},  \pg{873--894}.

\bibitem[Sarpkaya(2004)]{Sarpkaya2004}
{\sc \au{Sarpkaya, Turgut}} \yr{2004}  \at{A critical review of the intrinsic
  nature of vortex-induced vibrations}.  \jt{Journal of fluids and structures}
  \bvol{19}~(4),  \pg{389--447}.

\bibitem[Sfakiotakis {\em et~al.\/}(1999)Sfakiotakis, Lane \&
  Davies]{Sfakiotakis1999}
{\sc \au{Sfakiotakis, Michael}, \au{Lane, David~M} \& \au{Davies, J Bruce~C}}
  \yr{1999}  \at{Review of fish swimming modes for aquatic locomotion}.
  \jt{IEEE Journal of oceanic engineering}  \bvol{24}~(2),  \pg{237--252}.

\bibitem[Shelley \& Zhang(2011)]{Shelley2011}
{\sc \au{Shelley, Michael~J} \& \au{Zhang, Jun}} \yr{2011}  \at{Flapping and
  bending bodies interacting with fluid flows}.  \jt{Annual Review of Fluid
  Mechanics}  \bvol{43},  \pg{449--465}.

\bibitem[Shoele \& Mittal(2016)]{Shoele2016}
{\sc \au{Shoele, Kourosh} \& \au{Mittal, Rajat}} \yr{2016}  \at{Energy
  harvesting by flow-induced flutter in a simple model of an inverted
  piezoelectric flag}.  \jt{Journal of Fluid Mechanics}  \bvol{790},
  \pg{582--606}.

\bibitem[Silva-Leon {\em et~al.\/}(2019)Silva-Leon, Cioncolini, Nabawy, Revell
  \& Kennaugh]{Silva2019}
{\sc \au{Silva-Leon, Jorge}, \au{Cioncolini, Andrea}, \au{Nabawy, Mostafa~RA},
  \au{Revell, Alistair} \& \au{Kennaugh, Andrew}} \yr{2019}  \at{Simultaneous
  wind and solar energy harvesting with inverted flags}.  \jt{Applied Energy}
  \bvol{239},  \pg{846--858}.

\bibitem[Tang {\em et~al.\/}(2015)Tang, Liu \& Lu]{Tang2015}
{\sc \au{Tang, Chao}, \au{Liu, Nan-Sheng} \& \au{Lu, Xi-Yun}} \yr{2015}
  \at{Dynamics of an inverted flexible plate in a uniform flow}.  \jt{Physics
  of Fluids}  \bvol{27}~(7),  \pg{073601}.

\bibitem[Tavallaeinejad {\em et~al.\/}(2020{\natexlab{{\em
  a\/}}})Tavallaeinejad, Legrand \& Pa{\"\i}doussis]{Tavallaeinejad2020a}
{\sc \au{Tavallaeinejad, Mohammad}, \au{Legrand, Mathias} \&
  \au{Pa{\"\i}doussis, Michael~P}} \yr{2020{\natexlab{{\em a\/}}}}
  \at{Nonlinear dynamics of slender inverted flags in uniform steady flows}.
  \jt{Journal of Sound and Vibration}  \bvol{467},  \pg{115048}.

\bibitem[Tavallaeinejad {\em et~al.\/}(2020{\natexlab{{\em
  b\/}}})Tavallaeinejad, Pa{\"\i}doussis, Legrand \&
  Kheiri]{Tavallaeinejad2020b}
{\sc \au{Tavallaeinejad, Mohammad}, \au{Pa{\"\i}doussis, Michael~P},
  \au{Legrand, Mathias} \& \au{Kheiri, Mojtaba}} \yr{2020{\natexlab{{\em
  b\/}}}}  \at{Instability and the post-critical behaviour of two-dimensional
  inverted flags in axial flow}.  \jt{Journal of Fluid Mechanics}  \bvol{890}.

\bibitem[Taylor {\em et~al.\/}(2001)Taylor, Burns, Kammann, Powers \&
  Wel]{Taylor2001}
{\sc \au{Taylor, George~W}, \au{Burns, Joseph~R}, \au{Kammann, Sean~M},
  \au{Powers, William~B} \& \au{Wel, Thomas~R}} \yr{2001}  \at{The energy
  harvesting eel: a small subsurface ocean/river power generator}.  \jt{Oceanic
  Engineering, IEEE Journal of}  \bvol{26}~(4),  \pg{539--547}.

\bibitem[Vogel(1989)]{Vogel1989}
{\sc \au{Vogel, Steven}} \yr{1989}  \at{Drag and reconfiguration of broad
  leaves in high winds}.  \jt{Journal of Experimental Botany}  \bvol{40}~(8),
  \pg{941--948}.

\bibitem[Vogel(1994)]{Vogel1994}
{\sc \au{Vogel, Steven}} \yr{1994} {\em Life in moving fluids\/}.
  \publ{Princeton University Press}.

\bibitem[Williamson \& Govardhan(2004)]{Williamson2004}
{\sc \au{Williamson, CHK} \& \au{Govardhan, R}} \yr{2004}  \at{Vortex-induced
  vibrations}.  \jt{Annu. Rev. Fluid Mech.}  \bvol{36},  \pg{413--455}.

\bibitem[Yu {\em et~al.\/}(2017)Yu, Liu \& Chen]{Yu2017}
{\sc \au{Yu, Yuelong}, \au{Liu, Yingzheng} \& \au{Chen, Yujia}} \yr{2017}
  \at{Vortex dynamics behind a self-oscillating inverted flag placed in a
  channel flow: Time-resolved particle image velocimetry measurements}.
  \jt{Physics of Fluids}  \bvol{29}~(12),  \pg{125104}.

\bibitem[Zhang {\em et~al.\/}(2000)Zhang, Childress, Libchaber \&
  Shelley]{Zhang2000}
{\sc \au{Zhang, Jun}, \au{Childress, Stephen}, \au{Libchaber, Albert} \&
  \au{Shelley, Michael}} \yr{2000}  \at{Flexible filaments in a flowing soap
  film as a model for one-dimensional flags in a two-dimensional wind}.
  \jt{Nature}  \bvol{408}~(6814),  \pg{835--839}.

\end{thebibliography}

\end{document}